\documentclass[sigconf]{acmart}

\usepackage{acmart-taps}
\usepackage{enumitem}
\setcopyright{none}
\usepackage{amsmath}
\usepackage{booktabs}
\usepackage{longtable}
\usepackage{tabularx}
\usepackage{array}
\usepackage{colortbl}
\usepackage{adjustbox}
\usepackage{graphicx}
\usepackage{xurl}
\usepackage{float}
\usepackage{xcolor}
\usepackage{soul}
\sethlcolor{yellow}

\providecommand{\findingclaim}[1]{%
  \begingroup
    \sethlcolor{blue!8}%
    \textbf{\hl{#1}}%
  \endgroup
}

\begin{document}

\title{How Mental Health Self-Disclosure Becomes Visible: Evidence from Eight Conditions on Reddit}

%\acmSubmissionID{XXXX}

\author{Renkai Ma}
\authornote{Both authors contributed equally to this research.}
\orcid{0000-0002-4434-2235}
\email{mark@ucmail.uc.edu}
\affiliation{
  \department{School of Information Technology}
  \institution{University of Cincinnati}
  \city{Cincinnati}
  \state{Ohio}
  \country{USA}
}

\author{Lingyao Li}
\authornotemark[1]
\orcid{0000-0001-5888-8311}
\email{lingyaoli@arizona.edu}
\affiliation{
  \department{College of Information Science}
  \institution{University of Arizona}
  \city{Tucson}
  \state{Arizona}
  \country{USA}
}

\author{Shanting Chen}
\orcid{0000-0003-3390-2513}
\email{chenshanting@ufl.edu}
\affiliation{
  \department{Department of Psychology}
  \institution{University of Florida}
  \city{Gainesville}
  \state{Florida}
  \country{USA}
}

\author{Chen Chen}
\orcid{0000-0001-7179-0861}
\email{chechen@fiu.edu}
\affiliation{
  \department{Department of Computer Science}
  \institution{Florida International University}
  \city{Miami}
  \state{Florida}
  \country{USA}
}

\author{Fan Yang}
\email{YANG259@mailbox.sc.edu}
\affiliation{
  \department{School of Journalism and Mass Communications}
  \institution{University of South Carolina}
  \city{Columbia}
  \state{South Carolina}
  \country{USA}
}

\author{Yuanyuan Lei}
\orcid{0000-0002-9753-8071}
\email{yuanyuan.lei@ufl.edu}
\affiliation{
  \department{Department of Computer and Information Science and Engineering}
  \institution{University of Florida}
  \city{Gainesville}
  \state{Florida}
  \country{USA}
}

\renewcommand{\shortauthors}{Ma et al.}
\begin{abstract}
People share mental health diagnoses on social media, yet how such language becomes visible around their self-disclosure, and whether community engagement tracks it, remain unexamined across conditions. We analyze 89,605 Reddit posts from 739 users across eight conditions, removing each user's diagnosis disclosure and aligning their surrounding posts to that anchor. Within the pre-disclosure year, language-visible burden was highest in the month before disclosure for six conditions, earlier for post-traumatic stress disorder and furthest from it for borderline personality disorder, and remained visible afterward rather than resolving. The theme Seeking Clinical Explanations showed the largest early-to-late difference before disclosure in five conditions, yet engagement rarely tracked what users wrote: only 9 of 360 language--engagement correlations survived correction. Disclosure is therefore a waypoint in an unevenly visible process, and we offer implications for community practice and platform design where engagement metrics do not reflect clinical need.
\end{abstract}




\maketitle

\section{Introduction}
\label{sec:introduction}

Social media platforms are common settings in which people express distress, seek support, and attempt to make sense of their mental health experiences~\citep{berry2017whywetweetmh,dechoudhury2014reddit,pendse2023marginalization, skaik2020using}. When a Reddit user writes, ``\textit{I was finally diagnosed with depression},'' readers may interpret the post as the point at which the condition is first explicitly named. For the author, however, the disclosure may represent a midpoint in a longer process of experiencing symptoms and sharing them with communities. Traces of this process may already be visible in the author’s earlier posts~\citep{chaudoir2010disclosure}. Understanding these posting histories could help identify when distress and diagnosis-related sense-making begin to emerge, thereby supporting outreach or encouragement to seek professional care before a user explicitly names a condition as their own. Self-disclosure itself can foster self-validation and social connection, but it can also expose users to stigma and discrimination~\citep{andalibi2016sexualabuse,edwards2024autism}. These histories also let researchers assemble condition-specific cohorts from self-reported diagnoses and compare their language with that of other users~\citep{coppersmith2015adhd,cohan2018smhd,jiang2020detection}.

Prior research has drawn two main insights from these social media posts. First, mental health sensing and early-detection research has treated language surrounding self-disclosure as predictive input, asking whether social media text can be used to identify mental health conditions~\citep {guntuku2017detecting,chancellor2020methods,mansoor2024early, kim2020deep}. Second, research on online communities has examined how audiences respond to disclosure and the forms of social support that users subsequently receive~\citep{andalibi2017sensitive,sharma2018mental,ernala2018audience}. Less attention has been paid to the process that connects these two perspectives, namely, how expressions of distress and diagnosis-related sense-making change around disclosure, and how community attention reacts to those expressions. A small number of studies have begun to address this gap by using disclosure as a temporal anchor. For example, one traced Reddit posting histories surrounding depression diagnosis disclosures~\citep{biester2022identity}, and another traced language around self-harm disclosures among users who self-identified as having borderline personality disorder (BPD)~\citep{entwistle2025bpd}. This event-centered research remains limited, however, and prior studies have not compared patterns surrounding diagnosis disclosure across multiple common mental health conditions using a common design.

Understanding how language changes around diagnosis disclosure has implications for timely support. Knowing when and how distress signals appear could inform the allocation of resources that offer support or encourage earlier professional help-seeking in this process, without assuming that social media language can establish a diagnosis. This temporal context is also important for community members and platform stakeholders, who decide which posts warrant attention based on limited textual and community engagement. Posts expressing substantial stress or burden may receive little engagement, whereas highly engaged posts do not necessarily reflect greater clinical need. We therefore use \emph{language-visible burden} for the condition-relevant burden expressed in an author’s posts; the construct describes a property of the text, not clinical status or symptom severity. Likewise, community engagement metrics capture observable reactions on the platform rather than the quality of support an author receives. We draw on the \textit{Disclosure Processes Model} \citep{chaudoir2010disclosure} and CSCW research on \textit{networked visibility} \citep{barta2024visibility} to conceptualize mental health disclosure as a temporal and socially situated process. These perspectives motivate us to examine both authors' language and recorded community engagement across the periods surrounding disclosure on Reddit, a text-first platform with persistent posting histories. We ask three Research Questions (RQs), one for each dimension: \emph{temporal progression}, \emph{discussion content}, and \emph{community reactions}.

\begin{itemize}[leftmargin=*, nosep]
    \item \textbf{RQ1 (Temporal progression):} \textit{How does the intensity of language-visible burden vary before and after a mental health diagnosis self-disclosure on Reddit?}
    \item \textbf{RQ2 (Discussion content):} \textit{Which burden dimensions and thematic content characterize posts surrounding disclosure?}
    \item \textbf{RQ3 (Community reaction):} \textit{How does recorded community engagement co-occur with these language indicators?}
\end{itemize}

To answer these questions, we analyze 89,605 Reddit posts from 739 users across eight mental health conditions. Using conservative text-based criteria (see Section~\ref{sec:methods}), we select one first-person diagnosis disclosure per user and align the surrounding posts to it; we then assess indicators of language-visible burden, affect, and thematic content, along with associations between these indicators and Reddit engagement that survive correction for multiple comparisons. Our findings show that disclosure does not follow a single trajectory across conditions. Within the pre-disclosure year, language-visible burden peaks immediately before disclosure for six conditions, earlier for post-traumatic stress disorder, and furthest from disclosure for BPD. Burden language then remains visible across the post-disclosure year (RQ1). Condition-specific expressions vary across cohorts, but the \emph{Seeking Clinical Explanations} theme shows the largest difference between the early and late pre-disclosure periods for five conditions (RQ2). Only 9 of 360 associations between language indicators and community engagement meet the corrected threshold (RQ3): what users write about their difficulties rarely tracks the attention their posts receive. These findings motivate \textit{mental health self-disclosure visibility} as a conceptual lens for understanding how condition-related experiences become interpretable and socially attended to around a self-disclosure. The lens treats disclosure as an event around which users frame their experiences and networked publics respond. It also keeps textual indicators and community engagement metrics as bounded traces of platform activity rather than measures of clinical severity or received support.

Our study makes four contributions to HCI, social computing, and digital mental health. First, it provides a temporal, multi-condition analysis of how mental health language changes around a diagnosis self-disclosure, extending research beyond single-condition studies of depression diagnosis claims or self-harm disclosures. Second, it offers a measurement approach that keeps author-expressed distress and community engagement analytically distinct, reducing the risk of collapsing clinical status and platform activity into a single measure. Third, it shows that within these cohorts recorded engagement corresponds only sparsely with condition-relevant language, cautioning researchers and platform stakeholders against reading Reddit engagement metrics as a reliable indicator of clinical need. Fourth, it translates these patterns into implications for community practice and social media platform design.

\section{Related Work} 
\label{sec:related-work}
We situate our study in three areas: how users disclose mental health conditions on social media, how researchers interpret mental health language and platform engagement, and the two theoretical lenses that guide our investigation.

\subsection{Mental Health Self-Disclosure on Social Media}
\label{sec:self-disclosure-related-work}
Self-disclosure broadly refers to communicating personal information about oneself to others \citep{qian2007anonymity}. On social media, it often takes the form of a first-person statement in which an author identifies themselves as having a particular condition. Researchers have frequently used such statements to construct condition-based cohorts. On Twitter, \citet{coppersmith2015adhd} compared aggregate language patterns across ten cohorts identified through self-reported diagnosis statements; \citet{cohan2018smhd} adapted this approach to nine Reddit cohorts; and \citet{jiang2020detection} examined language patterns and classification performance across eight. These studies demonstrated the value of first-person disclosures for identifying condition-based cohorts, but their primary analyses did not distinguish posts written before disclosure from those written afterward, potentially obscuring temporal changes in how users express and interpret their mental health experiences. We next review self-disclosure research on the eight conditions in our study: depression, anxiety, ADHD, autism, PTSD, OCD, bipolar disorder, and BPD.

Prior work on \textbf{depression} comes closest to tracing language around a disclosure. An analysis of Instagram posts tagged \#depression found that the form of disclosure and the expression of support needs were associated with the comments a post received \citep{andalibi2017sensitive}. On Reddit, \citet{biester2022identity} centered authors' historical posts in which they had received a depression diagnosis, which the authors termed \emph{identity claims}. Language reflecting anxiety, sadness, and cognitive processing increased before these claims and then declined toward matched-control levels. This study provides the closest temporal precedent for our approach, but it focuses on one form of disclosure within a single condition.

Research on \textbf{anxiety} shows that mental health language is sensitive to its context. \citet{ireland2018within} found that the same Reddit users wrote differently in anxiety-support and neutral communities, and that members of anxiety communities differed from other users even when both wrote in neutral communities. \citet{li2024socialanxiety} found that Reddit and Douban social-anxiety posts differed in their common themes and language use. Neither study organizes an author's posting history around first-person condition self-disclosure.

Research on Attention-Deficit/Hyperactivity Disorder (\textbf{ADHD}) treats self-disclosure as diagnosis- and identity-related sensemaking rather than as the communication of medical information. \citet{eagle2023adhd} found that ADHD communities on TikTok, Twitter, and Instagram serve as sources of experiential knowledge and validation while surfacing tensions around diagnosis and medical authority. \citet{leveille2024adhd} found that TikTok self-disclosures framed through humor and personal experience function as identity work. Neither study traces how this process becomes visible in an individual author's posts before and after disclosure.

Prior work on \textbf{autism} shows that the meaning and consequences of disclosure depend on the surrounding social environment. \citet{edwards2024autism} analyzed 3,121 Reddit and Twitter posts and linked public misunderstanding with social stigma and with concerns about privacy and respectful treatment. \citet{romualdez2021workplace} found that autistic adults' disclosure outcomes varied across workplaces and organizational cultures. Because both studies focus on public discourse or retrospective accounts, neither examines how an individual author's language changes around a disclosure.

Research on Post-Traumatic Stress Disorder (\textbf{PTSD}) provides an early example of using self-reported diagnosis statements to construct a social media cohort. \citet{coppersmith2014ptsd} identified and manually reviewed tweets in which authors reported receiving a PTSD diagnosis, then compared those authors' timelines with randomly sampled users. Although the diagnosis statements established cohort membership, they were not used as temporal anchors for comparing the authors' posts before and after disclosure.

Research on Obsessive-Compulsive Disorder (\textbf{OCD}) has focused largely on participation in online peer-support communities. Peer-support users have reported misinformation and increased symptom preoccupation as negative experiences \citep{tan2021ocd}, and a thematic analysis of r/OCD found that users sought validation and connection by sharing symptom experiences, although some exchanges may also have functioned as reassurance-seeking \citep{sun2025ocd}. This work clarifies how OCD-related communication operates within peer-support settings, but it remains unclear whether practices such as symptom-sharing and reassurance-seeking change when an author explicitly discloses having OCD.

Prior work on \textbf{bipolar disorder} has examined community discourse and diagnosis-based cohorts cross-sectionally: medication received more attention in r/Bipolar and suicide more in r/Depression \citep{yoo2019bipolar}, and Reddit users who self-reported a bipolar diagnosis often reported additional mental health diagnoses \citep{jagfeld2021bipolar}. Whether such topics become more prominent as authors approach or move beyond a bipolar disorder disclosure remains unexamined.

Research on Borderline Personality Disorder (\textbf{BPD}) connects condition traits with posting and interaction behavior. In a nonclinical survey, participants with higher BPD-trait scores reported posting, regretting or deleting posts, and blocking others more often \citep{ooi2020bpd}. A recent Reddit study traced language around disclosures of suicidality and nonsuicidal self-injury among users who self-identified as having BPD and examined the scores and replies those posts received \citep{entwistle2025bpd}. Although the focal events were self-harm disclosures rather than BPD disclosures, this study provides an important temporal precedent for examining language and engagement around a disclosure event within a BPD cohort.

Prior work thus offers insights into condition-specific language, peer support, and audience response, but no study integrates these strands within a common temporal design. Multi-condition studies typically define cohorts without aligning posts around the disclosures, whereas temporal studies have focused on single conditions. Our study brings the two together, applying consistent self-disclosure criteria and fixed temporal windows across eight conditions to compare how authors' language changes before and after disclosure while preserving condition-specific patterns. Reddit suits this design because accounts have persistent, publicly retrievable posting histories that can be aligned with a disclosure event, and because its posts contain sufficient running text to support lexical measurement \citep{dechoudhury2014reddit}.

\subsection{Interpreting Mental Health Expressions and Engagement on Social Media} 
While social media offers rich data for examining mental health, prior work documents two primary challenges: defining what study cohorts actually represent, and interpreting the community engagement that surrounds them. First, computational models often define their study cohorts using self-reported disclosure labels (e.g., posts stating ``I was diagnosed with depression''), and reviews caution that these labels capture a specific social media action rather than a verified clinical diagnosis \citep{guntuku2017detecting, chancellor2020methods}. Treating these digital traces as direct stand-ins for clinical populations can lead to mismatched expectations, as \citet{ernala2019methodological} found that models trained on common social media proxies performed poorly when tested against data from clinically diagnosed patients. Similarly, \citet{chancellor2023contextual} demonstrated that relying solely on disclosure posts captures only a narrow, situated facet of the broader eating-disorder experience. This disconnect highlights the importance of construct validity: researchers must clearly separate the underlying concept they want to study from the digital traces they actually measure \citep{cronbach1955construct, jacobs2021measurement, messick1995validity}. We therefore do not attempt to predict ground-truth clinical status; we interpret lexical and topical patterns in user posts as textual expressions of distress rather than clinical measures; Section~\ref{subsec:burdendimension} defines this construct as \textit{language-visible burden}.

Second, the online community engagement surrounding user posts, such as likes and comments, presents its own interpretive challenges. That a researcher can access a post does not mean an audience saw or engaged with it \citep{boyd2010networkedpublics}. Early studies showed that users substantially underestimate their audience size and that raw engagement numbers are weak indicators of true reach \citep{bernstein2013invisible}. Platform mechanisms such as algorithmic recommendation further shape who sees mental health content and how users assess their own visibility \citep{barta2024visibility, milton2023tiktok}. Computational researchers cannot measure the empathy behind a like or a comment, so these metadata cannot stand in for it \citep{olteanu2019socialdata}. We therefore treat Reddit scores, upvote ratios, and comment counts as observable online community reactions that co-occur with users’ language, rather than as estimates of social support or clinical need \citep{barta2024visibility, salganik2017bitbybit}.

\subsection{Theoretical Lens: The Disclosure Process and Networked Visibility}
\label{sec:rw-theoretical-foundations}
Treating self-disclosure as a static label obscures how mental health experiences are expressed over time and encountered within online communities. We thus integrate two complementary theoretical perspectives: the \textit{Disclosure Processes Model}, which situates disclosure within an ongoing process, and networked visibility, which distinguishes what users express from how their content becomes observable socially.

 The \textit{Disclosure Processes Model} conceptualizes self-disclosure as more than an isolated act~\citep{chaudoir2010disclosure}. Developed to explain how people disclose concealable stigmatized identities, including mental illness, the model separates the goals that precede a disclosure from the processes that follow it. Approach or avoidance goals shape whether and how people disclose; the disclosure event can then trigger mediating processes, such as reduced inhibition, social support, and changes in what others know, whose outcomes feed back into future disclosure decisions \citep{chaudoir2010disclosure}. Prior HCI research similarly shows that people may spend extended periods of ``testing the waters'' before disclosing a stigmatized identity \citep{andalibi2018testing} and may use social media to negotiate condition-related identities \citep{eagle2023adhd, leveille2024adhd}.  From this process-oriented perspective, a first-person diagnosis disclosure on Reddit serves as a temporal anchor within a broader period of distress, uncertainty, and sensemaking rather than as the beginning of the author's mental health experience \citep{biester2022identity}. This perspective motivates our examination of how authors' language changes before and after disclosure.

Networked visibility complements this temporal perspective by explaining why what authors express should be distinguished from the attention their posts receive. CSCW scholarship conceptualizes visibility as a multidimensional affordance of social media: what an author posts, how it circulates through platform mechanisms, and the reactions it receives are related but distinct dimensions \citep{treem2013social}. Algorithms determine which content a platform amplifies and who sees it \citep{devito2017algorithms, milton2023tiktok}, while a post's potential audience may remain invisible to its author \citep{bernstein2013invisible}. The Visibility Objects lens further emphasizes that users may evaluate the visibility of particular content separately from the visibility of their broader identities \citep{barta2024visibility}. Together, these perspectives shift the focus from whether social media text can identify a diagnostic category to how condition-related experiences become linguistically visible around disclosure and whether recorded community attention aligns with those expressions.

\section{Methods}
\label{sec:methods}

\subsection{Collecting Mental Health Self-Disclosure Data from Reddit}
\label{sec:data-collection}
We constructed the corpus in three stages: identifying candidate disclosure authors during a predefined collection period, retrieving their available posting histories, and screening those histories to establish the initial cohort.

\textbf{Step 1. Identifying candidate disclosure authors.}
We used Brandwatch\footnote{\url{https://www.brandwatch.com/}} to search historical Reddit data because it provided comprehensive search coverage. Brandwatch also supported site-wide search, allowing us to identify posts across Reddit rather than restricting collection to a predefined set of communities. We searched for posts published in 2024 that contained both a first-person disclosure phrase and a mental health term from the query list in Table~\ref{tab:search-terms}. The query included 41 first-person disclosure phrases, such as ``was diagnosed with'' and ``my psychiatrist said I have.'' The mental health terms included both general phrases, such as ``mental health'' and ``mental illness,'' and terms for conditions beyond the eight examined in this study. Because Brandwatch returned Reddit post identifiers but not the full post records required for analysis, we used Reddit's official API via the Python Reddit API Wrapper (PRAW)\footnote{\url{https://praw.readthedocs.io/en/stable/}} to retrieve the corresponding content and metadata.

We limited the Brandwatch search to 2024 because history retrieval ended in early 2026, providing approximately one to two years in which to observe platform activity after each matched post. A 2025 collection period would have provided substantially less follow-up time. The 2024 restriction applied only to the posts used to identify candidate authors; candidate disclosures subsequently identified within an author's retrieved history could occur outside 2024. After deduplicating author accounts, the search yielded 83,140 candidate disclosure authors.

\textbf{Step 2. Retrieving author histories.}
For each of the 83,140 authors, we used PRAW to retrieve all posts, yielding a total of 6,247,159 posts. For each post, we recorded the identifier, title, body text, creation timestamp, subreddit, permalink, author name, post-type and content flags, and engagement fields. The analyses use three engagement measures: score, upvote ratio, and comment count. Although author identification was restricted to posts matched in 2024, the retrieved histories covered a broader period spanning August 2011 to February 2026.

\textbf{Step 3. Screening histories and establishing the cohort.}
We applied a set of rules to construct the cohort. First, we retained authors with at least 20 posts in their retrieved histories to ensure a level of observable posting activity for longitudinal analysis; minimum-activity filters of this kind are common in social media mental health research \citep{chancellor2020methods,cohan2018smhd}. Second, we screened these histories for posts in which a first-person diagnosis pattern and a condition term from Table~\ref{tab:search-terms} occurred together. Unlike the broader Brandwatch search, this screening applied fixed, case-insensitive patterns and required both term types to occur within the same post. Applying these two rules identified 12,724 candidate disclosure posts from 10,088 authors. We then applied strict first-person disclosure criteria to these posts, requiring an explicit self-attribution of the condition rather than a mention of it. Appendix~\ref{app:disclosure-anchor-criteria} states this requirement formally as $D_{i,c}$ and lists the wording that disqualifies a post, including uncertainty, denial, quoted or relayed disclosures, and symptom-only descriptions. The single-condition and uncertainty gates that accompany $D_{i,c}$ in Eq.~\eqref{eq:candidate-disclosure-gate} are applied later, in Section~\ref{sec:cohort-construction}. This process retained 4,679 strict disclosure posts.

For authors with multiple retained disclosures, we designated the earliest qualifying post as the \emph{initial candidate disclosure}, yielding 3,784 unique authors. Because the broad query captured disclosures of many conditions, we restricted the analysis to the eight most frequently represented conditions, each of which supplied enough authors to estimate condition-level means within every event-time window: ADHD, anxiety, autism, bipolar disorder, BPD, depression, OCD, and PTSD. Selecting conditions by the volume of available self-reported disclosures is established practice in this literature: prior multi-condition studies analyze ten \citep{coppersmith2015adhd}, nine \citep{cohan2018smhd}, and eight \citep{jiang2020detection} conditions chosen the same way, and report that the conditions with the fewest users yield the least reliable estimates \citep{cohan2018smhd}. Each condition we retain appears in at least one of those sets. We further required each retrieved history to contain the initial candidate disclosure and at least five posts preceding it. These criteria yielded an initial cohort of 1,881 authors, each assigned one study condition and one initial candidate disclosure. Section~\ref{sec:cohort-construction} describes the subsequent selection of disclosure anchors and construction of the final analytic cohort.

\subsection{Disclosure Cohort Construction}
\label{sec:cohort-construction}
\begin{figure*}[t]
    \centering
    \includegraphics[width=0.9\textwidth]{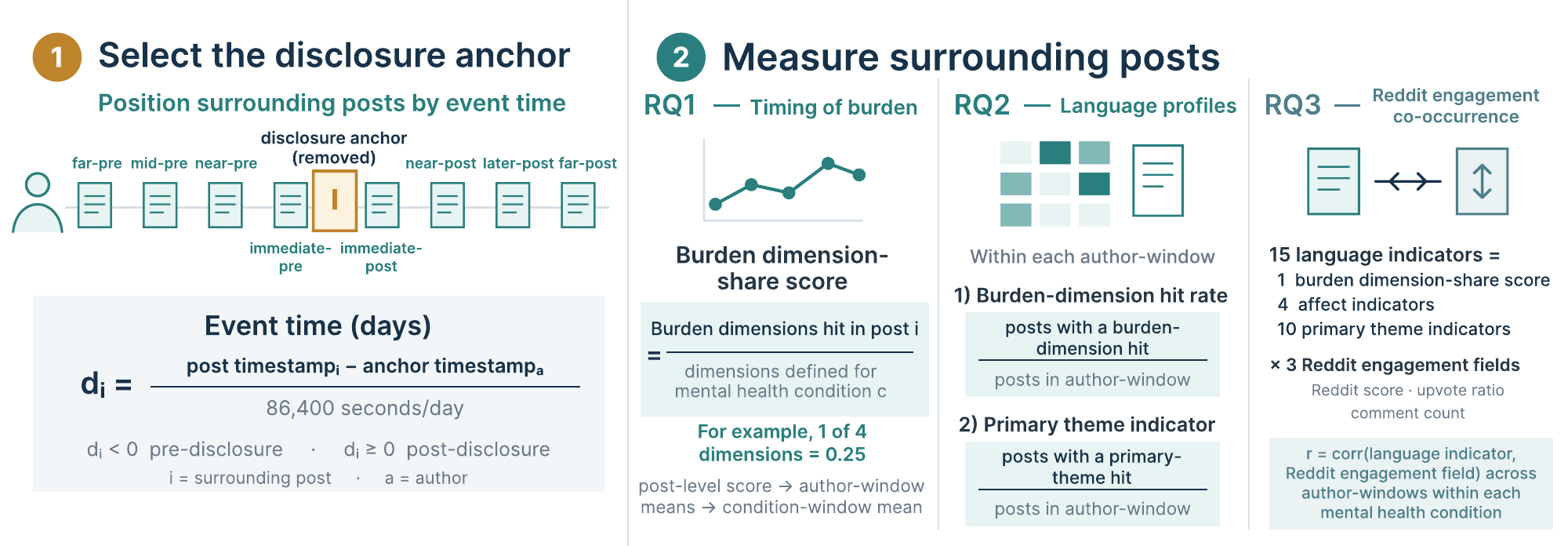}
    \caption{Analysis overview: one disclosure anchor is selected and removed per author, surrounding posts are positioned in eight event-time windows, post-level measures are aggregated into author-window and condition-window summaries (RQ1--RQ2), and RQ3 correlates 15 language indicators with three Reddit engagement fields in the same author-window.}
    \Description{Two-stage overview of data analysis. The first stage shows an author's surrounding posts positioned in eight event-time windows around a removed disclosure anchor and presents the event-time equation. The second stage maps RQ1 to the burden dimension-share score, RQ2 to burden-dimension and primary-theme measures, and RQ3 to same-window correlations between language indicators and Reddit engagement fields.}
    \label{fig:methods-overview}
\end{figure*}

We use a descriptive event-time design centered on one selected first-person diagnosis disclosure per author. The selected post, which explicitly links one of the eight study conditions to the author, serves as the \emph{disclosure anchor}. We remove this post and analyze the author’s remaining \emph{surrounding posts}. The anchor is treated only as a temporal reference point, not as the author’s first disclosure, diagnosis date, condition onset, or a causal event; accordingly, pre-disclosure and post-disclosure refer only to time relative to this anchor. Figure~\ref{fig:methods-overview} summarizes the analysis. We position surrounding posts by event time, derive measures of language-visible burden, affect, and thematic content, and aggregate them by author-window and condition for RQ1–RQ2. For RQ3, we examine within-condition associations between these language indicators and three Reddit engagement fields. Appendix~\ref{app:measurement} provides the corresponding equations.

\subsubsection{Selecting First-Person Disclosure Anchors}
For each of the 1,881 authors retained in Section~\ref{sec:data-collection}, we used a two-stage procedure to determine whether the initial candidate or an earlier post should serve as the disclosure anchor; Appendix~\ref{app:disclosure-anchor-criteria} formalizes the eligibility criteria, condition-match score, and selection rule.
First, we excluded 1,016 authors whose initial candidate did not support an unambiguous single-condition assignment: 977 mentioned multiple study conditions and 39 expressed uncertainty about the assigned condition. We did not search for replacement anchors for these authors.
Second, for each remaining author, we evaluated the initial candidate and all earlier available posts. An \emph{eligible disclosure} had to name the assigned condition, mention no other study condition, and explicitly connect the condition to the author through first-person wording such as \emph{I have}, \emph{I am}, or \emph{I was diagnosed with}. These conservative criteria prioritize condition-specific anchors over cohort size.

For each evaluated post, we also recorded a \emph{condition-match score} as an audit field; it did not determine eligibility and only ordered eligible posts sharing the earliest timestamp. Because these fixed rules assign labels without manual review of every post, they constitute rule-based \emph{weak supervision} \citep{grimmer2013text,ratner2017snorkel}. Eligibility therefore indicates only that a post meets our textual disclosure criteria, not that the diagnosis or clinical status is verified \citep{chancellor2020methods,guntuku2017detecting}. We selected the earliest eligible post at or before the initial candidate as the disclosure anchor. Another 126 authors had no eligible post, leaving 739 authors; for 241, the selected anchor preceded the initial candidate. Because moving the anchor earlier reclassifies posts that preceded the initial candidate as post-disclosure, some retained authors have fewer than five pre-anchor posts even though every author met the five-post requirement relative to the initial candidate. We then reconstructed each retained author's available posting history and removed only the selected anchor, leaving other diagnosis-related posts in the surrounding-post corpus. The selected anchors were posted between May 1, 2013 and January 31, 2025. The final cohort contains 739 authors and 89,605 surrounding posts: 43,098 before and 46,507 after the anchor. Authors contribute only to event-time windows in which they have observed posts.

\textbf{Positioning Surrounding Posts Relative to the Disclosure Anchor.} For each surrounding post, we define \emph{event time} as the number of days from the author's disclosure anchor (Appendix~\ref{app:event-time-indexing}); negative values indicate pre-disclosure posts and values of zero or greater indicate post-disclosure posts. We divide the year before and after the anchor into eight windows: far-pre ($[-365,-180)$), mid-pre ($[-180,-90)$), near-pre ($[-90,-30)$), immediate-pre ($[-30,0)$), immediate-post ($[0,30)$), near-post ($[30,90)$), later-post ($[90,180)$), and far-post ($[180,365)$). These windows contain 44,773 posts from 728 authors, yielding 4,046 author-window observations; posts outside the 730-day span are retained in the full histories but excluded from event-time analyses and figures.

\subsection{Measuring Language in Surrounding Posts}
\label{sec:measuring-language}
\label{subsec:burdendimension}
We derive three classes of language indicators from surrounding posts: language-visible burden, affect indicators, and primary themes indicators. These measures describe observable text patterns rather than diagnosis, symptom severity, emotional state, or clinical need.

\textbf{(1) Language-visible burden.}
We use \emph{language-visible burden} as an umbrella term for condition-relevant difficulty language and the \emph{burden dimension-share score} as its primary post-level measure. Prior computational research has analyzed such language patterns while cautioning that they do not establish clinical status \citep{chancellor2020methods,ernala2019methodological}.

For each mental health condition, we define a literature-informed set of \emph{burden dimensions}, each representing a distinct type of condition-relevant difficulty that may appear in a post. We apply only the dimensions associated with the condition named in the author's disclosure anchor. For example, when an anchor names depression, we search the author's surrounding posts only for depression-related dimensions. Depression has five dimensions, whereas each other condition has four; Appendix Table~\ref{tab:burden-dimensions} lists their definitions and sources. For each dimension, we define search words and phrases and assign a \emph{burden-dimension hit} when at least one appears in a post, regardless of capitalization. Because this procedure matches text strings without interpreting sentence context, a hit may not always reflect the intended meaning. For example, a general reference to insomnia may match a term intended to capture decreased need for sleep in bipolar disorder. Section~3.5 evaluates sampled hits in context through human validation.

For each post, the \emph{burden dimension-share score} is the number of dimensions with at least one hit divided by the total number defined for that condition. For a four-dimension condition such as autism, a score of 0.25 indicates that one dimension was matched. As a secondary measure, we calculate a \emph{burden phrase rate}, which counts all matched burden phrases, including repetitions, per 100 word-like tokens. Appendix~\ref{app:burden-affect-implementation} defines both measures.

\textbf{(2) Affect indicators.}
We use \emph{affect indicators} to characterize emotional language expressed in each post. Unlike the condition-specific burden dimensions, these indicators use the same lexical categories across all eight conditions. Following lexicon-based sentiment and emotion methods \citep{hutto2014vader,mohammad2013nrc,pennebaker2015liwc}, we constructed study-specific word and phrase lists and counted case-insensitive matches in each post. For each category, we divide the number of matches by the number of word-like tokens and multiply by 100 to obtain an \emph{affect rate}.

The analyses use four affect indicators: negative-affect rate, positive-affect rate, \emph{sentiment balance}, and \emph{affective arousal}. Sentiment balance is calculated as the positive-affect rate minus the negative-affect rate, and affective arousal as the fear rate plus the arousal-intensity rate. Appendix~\ref{app:burden-affect-implementation}, Eqs.~\eqref{eq:affect-count}--\eqref{eq:affective-arousal}, provides the corresponding formulas.

\textbf{(3) Primary-theme indicators.} We use \emph{primary-theme indicators} to capture recurring ways authors frame and interpret their experiences in surrounding posts. To develop the codebook, one research team member randomly sampled 200 posts from the original surrounding-post corpus and open-coded them before application of the final disclosure-anchor criteria. This process produced 80 initial codes representing recurring ideas in authors' descriptions of their experiences. After team discussion, the first author compared and grouped related codes into ten \emph{primary themes}. For each theme, the codebook specifies its definition, boundaries, subthemes, and search terms \citep{braun2006thematic,nelson2020cgt}. Appendix~\ref{app:theme-codebook} presents the full codebook and short representative quotations.

We then fixed the codebook and applied its search terms to all 89,605 surrounding posts in the final cohort. A post receives a \emph{primary-theme hit} when at least one corresponding search term appears; because themes are not mutually exclusive, a post may receive multiple hits. We refer to themes by the names in Appendix Table~\ref{tab:theme-codebook}, with abbreviated labels in figures. The \emph{Imagining Self-Removal as an Escape From Distress} theme refers to withdrawal or self-erasure language; \emph{removal} here is unrelated to removing disclosure anchors from the timeline. For each theme, we average post-level hits within each author and event-time window to obtain an author-window hit rate, which we call the \emph{primary-theme indicator}. These indicators reflect automated application of the codebook rather than manual coding of every post. Appendix~\ref{app:primary-theme-labeling} specifies the automated assignment rule.

\subsection{Measuring Reddit Engagement Fields}
We analyze three non-duplicate Reddit engagement fields: \emph{Reddit score} (platform-reported voting score), \emph{upvote ratio} (share of recorded votes that are upvotes), and \emph{comment count} (number of comments). We exclude \texttt{response\_count} because it duplicates comment count. Throughout the paper, \emph{Reddit engagement} refers only to these metadata, which capture observable platform reactions but not response content or supportiveness, non-interacting viewers, or support quality \citep{andalibi2017sensitive,dechoudhury2014reddit}. We convert all three fields to numeric values and replace missing or unconvertible entries with zero, so zero may indicate either a recorded value or an imputed missing/invalid value \citep{jacobs2021measurement,messick1995validity,salganik2017bitbybit}. %We therefore treat these fields as bounded indicators of platform engagement rather than direct measures of audience exposure, social support, or clinical need.

\subsection{Human Validation}
Because fixed lexical patterns may match terms out of context or miss relevant expressions, we conduct human validation of automated burden-dimension and primary-theme assignments. Before applying the final disclosure-anchor criteria, we sampled 300 posts for burden validation and 300 post--theme cases for primary-theme validation. Because validation targets post-level lexical assignment rather than anchor selection or event-time positioning, we retain these samples. Of these cases, 111 burden posts and 106 theme cases remain in the surrounding-post timeline; neither sample contains a selected disclosure anchor, and cases outside the final timeline do not contribute to longitudinal estimates. The burden sample is stratified by condition and automated hit status, while the theme sample is evenly distributed across the ten themes and between automated hits and non-hits. Because of this stratified design, the samples do not estimate the prevalence of burden-dimension or primary-theme language in the full corpus or event-time distribution.

Coders worked in pairs and independently labeled each burden dimension or post--theme case as present, absent, or uncertain, without seeing the automated label. The 300 burden posts yielded 1,238 possible judgments because depression has five dimensions and each other condition four; the 300 theme cases yielded one judgment each. All three coder pairs completed both tasks, and Table~\ref{tab:validation} reports the 1,233 burden judgments and 297 theme cases completed by both coders. Burden-dimension hits were more often confirmed by human coders, whereas primary-theme indicators recovered more coder-identified language. We therefore interpret automated hits as evidence that corresponding language is explicit in a post, not that an underlying psychological or clinical state is present. Affect indicators were not separately validated and are likewise interpreted only as study-specific lexical measures of expressed affect.

\begin{table*}[t]
\centering
\footnotesize
\setlength{\tabcolsep}{5pt}
\renewcommand{\arraystretch}{1.05}
\caption[Human Validation]{Human Validation: Coder Agreement and Automated-Label Performance. \emph{Note.} Agreement and $\kappa$ describe the independent judgments, with uncertain judgments treated as absent. Performance is computed on the units where the two coders reached the same judgment. Because both samples are stratified by automated hit status, precision is the quantity this design estimates, and accuracy, recall, and F1 describe the sampled units only.}
\label{tab:validation}
\begin{adjustbox}{max width=\textwidth}
\begin{tabular}{@{}l r r r r r r r r@{}}
\toprule
& \multicolumn{3}{c}{\textbf{Coder agreement}} & \multicolumn{5}{c}{\textbf{Automated labels on agreed units}} \\
\cmidrule(lr){2-4} \cmidrule(lr){5-9}
\textbf{Task} & \textbf{Judgments} & \textbf{Agreement} & \textbf{$\kappa$} & \textbf{Units} & \textbf{Accuracy} & \textbf{Precision} & \textbf{Recall} & \textbf{F1} \\
\midrule
Burden dimensions & 1,233 & 89.3\% x& .74 & 1,101 & .79 & .72 & .37 & .49 \\
Primary themes & 297 & 87.2\% & .72 & 259 & .64 & .48 & .70 & .57 \\
\bottomrule
\end{tabular}
\end{adjustbox}
\end{table*}

\subsection{Statistical Analysis}
\textbf{Author-window aggregation.} Because authors contribute different numbers of posts, we aggregate each measure in two stages: first averaging post-level values within each author and event-time window to obtain an \emph{author-window value}, then averaging these values within each condition and window to obtain a \emph{condition-window mean}. Approximate 95\% confidence intervals are calculated as the condition-window mean ± 1.96 times the standard error across contributing author-window values. The same two stages produce \emph{author-window hit rates} for each burden dimension and primary theme. Appendix~\ref{app:author-window-estimates} provides the aggregation and confidence-interval equations.

\textbf{Identifying burden maxima (RQ1).} For each condition, we define the \emph{pre-disclosure maximum} as the pre-disclosure window with the largest condition-window mean burden dimension-share score and identify the maximum across all eight windows to characterize the post-disclosure trajectory. We conduct one paired \(t\)-test per condition among authors observed in both the far-pre and immediate-pre windows, and a parallel paired \(t\)-test among authors observed in both the immediate-pre and immediate-post windows. We report all sixteen paired tests without adjusting their $p$-values for multiple comparisons. Appendix~\ref{app:paired-cooccurrence-lagged} defines the paired differences.

\textbf{Checking burden-score sensitivity (RQ1).} We compare trajectories of the burden dimension-share score with the burden phrase rate defined in Section~\ref{sec:measuring-language}. This assesses whether temporal conclusions depend on the scoring choice.

\textbf{Summarizing disclosure language profiles (RQ2).} For each condition and burden dimension, the \emph{late-pre mean} pools author-window hit rates from the near-pre and immediate-pre windows. For each primary theme, the \emph{early-pre mean} pools far-pre and mid-pre rates, while the \emph{late-pre mean} pools near-pre and immediate-pre rates; the late-minus-early difference is their difference. Each \emph{pooled mean} averages all contributing author-window values across the two windows, so an author observed in both windows contributes two values; see Appendix~\ref{app:author-window-estimates}, Eq.~\eqref{eq:pooled-mean}. Because the two pools may contain different authors, these values compare pooled means rather than within-author longitudinal change.

\textbf{Testing Reddit engagement co-occurrence (RQ3).} Within each condition, we calculate Pearson correlations between 15 language indicators, including the burden dimension-share score, four affect indicators, and ten primary-theme indicators, and three Reddit engagement fields measured in the same author-window. These correlations describe co-occurrence, not causality or support received. We report \emph{presence share}, the proportion of author-windows in which an indicator exceeds zero; for sentiment balance, it is the proportion with a positive balance. For each language--engagement pair, we use a Mann--Whitney test to compare engagement between author-windows above zero and those at or below zero, with Pearson correlations as the primary analysis because zero differs in meaning across indicators. Across eight conditions, 15 indicators, and three engagement fields, this yields 360 Pearson correlations and 360 Mann--Whitney tests. We apply Benjamini--Hochberg (BH) correction separately to each test family to control the false discovery rate (FDR) at $q \leq .05$, and report $q$-values.

\textbf{Author-level sensitivity analysis.} Because authors may contribute multiple windows, we average each author's language and engagement values across observed windows, yielding one row per author. Testing the same 15 indicators against three engagement fields again produces 360 planned comparisons per statistic. All 360 Pearson correlations are estimable; 324 Mann--Whitney tests have sufficient observations, and BH correction is applied to those 324 finite $p$-values. This analysis assesses whether association directions persist after collapsing repeated author-window observations, rather than treating same-window $q$-values as definitive hypothesis tests.

\textbf{Exploratory lagged analysis.} We pair each author's language value in one window with engagement in the next adjacent window, retaining pairs only when both are observed. The analysis includes nine language indicators: the burden dimension-share score; negative affect; positive affect; affective arousal; and the themes \emph{Seeking Clinical Explanations}, \emph{Feeling Too Exhausted for Daily Functioning}, \emph{Trying to Eliminate Uncertainty Through Checking and Reassurance Seeking}, \emph{Losing Control of Sleep, Energy, Mood, and Thought Speed}, and \emph{Imagining Self-Removal as an Escape From Distress}. Across eight conditions and three engagement fields, this yields 216 Pearson correlations and 216 Mann--Whitney tests, with BH correction applied separately to each family. These analyses include no covariates and are exploratory temporal associations, not evidence of prediction or causality. Appendix~\ref{app:paired-cooccurrence-lagged} provides the same-window, author-level, and lagged equations, and  Appendix~\ref{app:bh-adjustment} defines the BH adjustment.

\section{Results}
\label{sec:results}

\subsection{RQ1 (Timing): Intensity of Language-Visible Burden Before and After Disclosure}
\label{sec:results-rq1}

\begin{figure*}[t]
\centering
\includegraphics[width=0.67\textwidth]{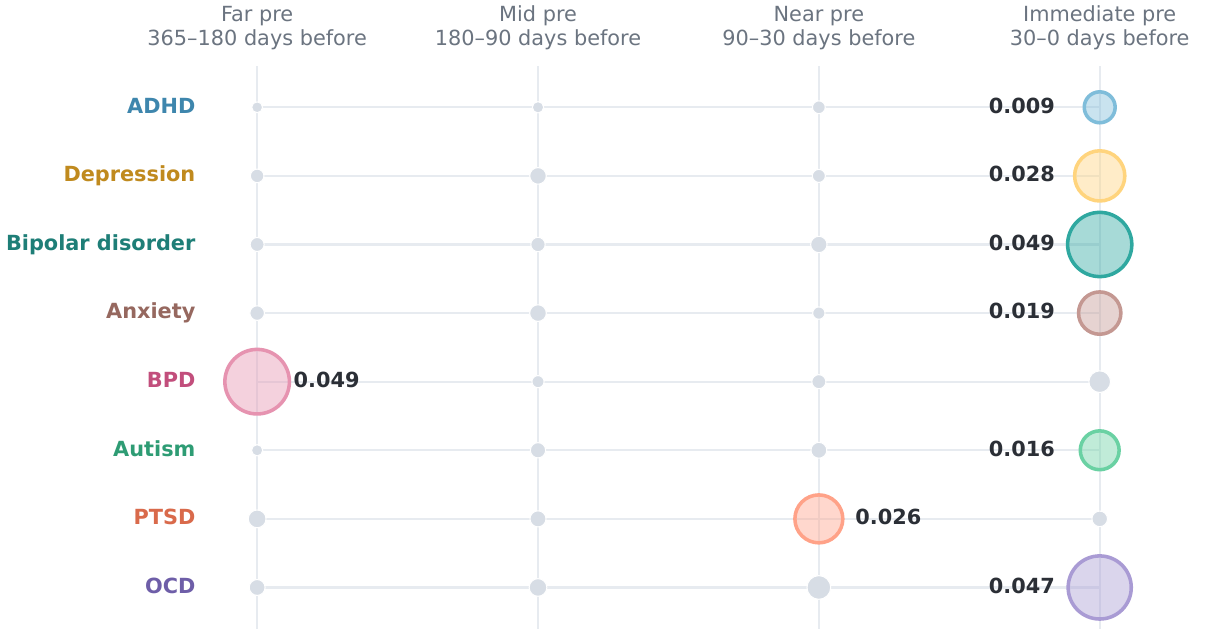}
\caption{Pre-disclosure peak timing by condition. The filled bubble marks the window with the highest condition-window mean of the burden dimension-share score, and bubble size indicates that mean; pale gray dots show the other pre-disclosure windows. Figure~\ref{fig:burden-heatmap} shows all eight windows.}
\Description{Peak timing map showing four pre-disclosure windows for eight mental health conditions. A filled colored bubble marks the highest condition-window mean of the burden dimension-share score for each condition, while pale gray dots show the other pre-disclosure windows. The far-pre window contains the highest condition-window mean for BPD, the near-pre window contains the highest mean for PTSD, and the immediate-pre window contains the highest mean for the other six conditions.}
\label{fig:peak-timing}
\end{figure*}

\findingclaim{Pre-disclosure burden peaks in the immediate-pre window for six conditions, earlier for PTSD, and furthest from disclosure for BPD.}
The largest condition-window mean falls in the immediate-pre window for ADHD, autism, depression, OCD, bipolar disorder, and anxiety, in the near-pre window for PTSD, and in the far-pre window for BPD (Figure~\ref{fig:peak-timing}). Language-visible burden therefore does not escalate uniformly toward disclosure.

Peak timing is sensitive to the burden operationalization. The burden phrase rate identifies the same pre-disclosure maximum as the burden dimension-share score for autism, depression, OCD, PTSD, bipolar disorder, and anxiety. For ADHD, the phrase-rate maximum occurs in the near-pre window rather than the immediate-pre window; for BPD, it occurs in the immediate-pre window rather than the far-pre window. The direction of the immediate-pre-minus-far-pre contrast agrees across the two measures for seven conditions and differs for BPD. The two operationalizations therefore support directional contrasts more consistently than identical peak locations.

\findingclaim{The magnitude and temporal variation of pre-disclosure burden differ across conditions.}
These descriptive maxima range from 0.009 for ADHD to 0.049 for bipolar disorder and BPD; Table~\ref{tab:1} reports all eight, and bubble size in Figure~\ref{fig:peak-timing} reflects each maximum. For BPD, for example, authors' within-window burden dimension-share scores average 4.9\% in the far-pre window: in an average post, 4.9\% of the BPD-specific burden dimensions have at least one lexical hit.

The pre-disclosure range---the difference between the lowest and highest pre-disclosure condition-window means---spans 0.045 for BPD down to 0.007 for ADHD (Table~\ref{tab:1}). Conditions with maxima in the same window can therefore still differ in how much their burden language varies: OCD and ADHD both peak in the immediate-pre window, but their ranges are 0.033 and 0.007. The author-windows contributing to each pre-disclosure maximum range from 249 (ADHD) to 11 (BPD); the BPD and anxiety maxima rest on small groups and should be read as descriptive summaries.

\findingclaim{Within-author comparisons do not support a common increase in burden as disclosure approaches.}
Within each condition, a paired \(t\)-test compares each author's mean burden dimension-share score in the far-pre window with that author's mean score in the immediate-pre window. The test evaluates whether the average within-author difference between the two windows differs from zero, rather than comparing the descriptive condition-window maxima. None of the eight unadjusted paired tests has a $p$-value below .05. The smallest unadjusted $p$-value occurs for ADHD ($n = 168$, mean immediate-pre-minus-far-pre difference = 0.007, $p = .053$). Taken together, the descriptive maxima and the paired tests indicate that language-visible burden does not follow a single shared trajectory toward disclosure. Table~\ref{tab:1} summarizes each condition's pre-disclosure maximum, leading late-pre burden dimensions, and same-window Pearson correlations that meet the BH-adjusted threshold.

\findingclaim{Burden language remains visible after disclosure, and four conditions reach their highest condition-window mean in a post-disclosure window.}
Across all eight event-time windows, those four maxima fall in the immediate-post window for ADHD (0.011) and autism (0.027), and in the near-post window for bipolar disorder (0.060) and anxiety (0.027). The other four conditions keep their pre-disclosure maxima (immediate-pre for depression and OCD, near-pre for PTSD, and far-pre for BPD). The within-author comparisons mirror this persistence: none of the eight unadjusted paired \(t\)-tests comparing each author's mean burden dimension-share score in the immediate-pre and immediate-post windows has a $p$-value below .05 (paired authors range from 16 for anxiety and BPD to 196 for ADHD). In this cohort, language-visible burden therefore persists into the post-disclosure year rather than resolving at the disclosure post; Figure~\ref{fig:burden-heatmap} displays the full eight-window profiles.

\begin{table*}[t]
\centering
\scriptsize
\setlength{\tabcolsep}{1.0pt}
\renewcommand{\arraystretch}{1.12}
\setlength{\emergencystretch}{2em}

\caption[Condition Profiles Around Self-Disclosure]{Condition Profiles Around Self-Disclosure. \emph{Note.} \texttt{Authors} counts all retained authors per condition ($N = 739$); the window-based columns draw on the 728 authors with posts inside the eight event-time windows. The score and range columns report the burden dimension-share score. \texttt{Leading late-pre burden dimensions} reports the two largest pooled means across the near-pre and immediate-pre windows. An em dash indicates that no Pearson correlation meets the BH-adjusted threshold.}
\label{tab:1}

\begin{tabular}{@{}>{\raggedright\arraybackslash}p{0.09\textwidth}
>{\raggedright\arraybackslash}p{0.068\textwidth}
>{\raggedright\arraybackslash}p{0.1\textwidth}
>{\raggedright\arraybackslash}p{0.075\textwidth}
>{\raggedright\arraybackslash}p{0.083\textwidth}
>{\raggedright\arraybackslash}p{0.2\textwidth}
>{\raggedright\arraybackslash}p{0.35\textwidth}@{}}
\toprule
\textbf{Condition} & \textbf{Authors} & \textbf{Pre-disclosure maximum window} & \textbf{Mean score at that window} & \textbf{Pre-disclosure range} & \textbf{Leading late-pre burden dimensions} & \textbf{Language--engagement correlation summary} \\
\midrule
ADHD & 389 & Immediate-pre & 0.009 & 0.007 & Organizational Lapses (0.013); Hyperactive Impulsivity (0.008) & 5 Pearson correlations meet the BH-adjusted threshold; the largest absolute correlation is the \emph{Imagining Self-Removal as an Escape From Distress} indicator with comment count ($r = .373$, $q < .001$). \\

Autism & 79 & Immediate-pre & 0.016 & 0.013 & Social Camouflaging (0.020); Sensory Overload (0.016) & --- \\

Depression & 63 & Immediate-pre & 0.028 & 0.022 & Low Mood (0.023); Self-Devaluation (0.023) & --- \\

OCD & 55 & Immediate-pre & 0.047 & 0.033 & Reassurance Seeking (0.064); Distressing Interference (0.049) & --- \\

PTSD & 61 & Near-pre & 0.026 & 0.013 & Involuntary Reexperiencing (0.028); Avoidant Shutdown (0.026) & --- \\

Bipolar disorder & 39 & Immediate-pre & 0.049 & 0.040 & Impulsive Risk and Crash (0.078); Manic Activation (0.048) & --- \\

Anxiety & 30 & Immediate-pre & 0.019 & 0.014 & Uncontrollable Worry (0.026); Acute Bodily Alarm (0.011) & --- \\

BPD & 23 & Far-pre & 0.049 & 0.045 & Abandonment Fear (0.044); Unstable Self-Experience (0.020) & 4 Pearson correlations meet the BH-adjusted threshold; the largest absolute correlation is negative-affect rate with comment count ($r = .541$, $q < .001$). \\
\bottomrule
\end{tabular}
\end{table*}

\subsection{RQ2 (Content): Burden Dimensions and Primary Themes in Surrounding Posts} \label{sec:results-rq2}
While RQ1 maps the temporal peaks of overall burden, RQ2 unpacks its content: the condition-specific difficulty language captured by the burden dimensions (Section~\ref{subsec:burdendimension}) and the primary themes that authors articulate.

\begin{figure*}[t]
\centering
\includegraphics[width=0.7\textwidth]{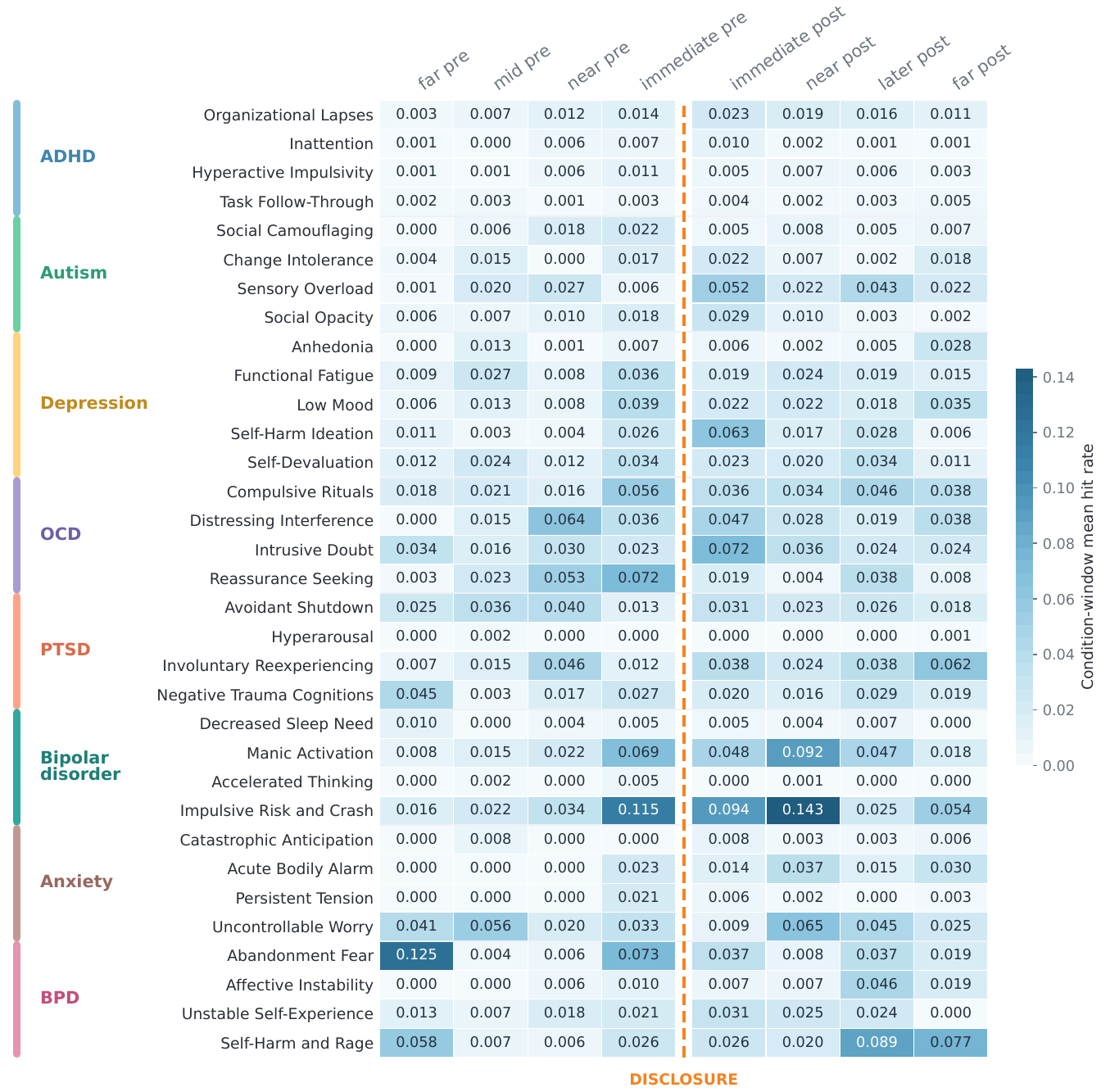}
\caption{Burden-dimension hit rates by event-time window. Each cell is the condition-window mean of author-level post hit rates for one condition-specific dimension.}
\Description{Heatmap of condition-window means for condition-specific burden-dimension hit rates across event-time windows. Rows are grouped by condition, and each cell averages authors' within-window post hit rates.}
\label{fig:burden-heatmap}
\end{figure*}

\findingclaim{The burden dimensions with the highest late-pre hit rates differ across conditions.} The burden dimension-share score in RQ1 combines condition-specific dimensions with different substantive content. Figure~\ref{fig:burden-heatmap} unpacks this composite score, showing that language-visible burden takes markedly different forms across the cohorts. Table~\ref{tab:1} reports the two largest pooled late-pre means for each condition. The leading dimension differs for every condition: Organizational Lapses for ADHD (0.013), Social Camouflaging for autism (0.020), Low Mood for depression (0.023), Reassurance Seeking for OCD (0.064), Involuntary Reexperiencing for PTSD (0.028), Impulsive Risk and Crash for bipolar disorder (0.078), Uncontrollable Worry for anxiety (0.026), and Abandonment Fear for BPD (0.044). Because the leading dimensions differ across conditions, the posts surrounding disclosure express condition-specific difficulties rather than only generic distress language.

\findingclaim{Several burden dimensions reach their highest hit rates in post-disclosure windows.}
The post-disclosure columns of Figure~\ref{fig:burden-heatmap} extend these condition-specific profiles beyond the disclosure anchor. For depression, the self-harm-ideation hit rate is higher in the immediate-post window (0.063) than in any pre-disclosure window (at most 0.026, in the immediate-pre window). The same pattern holds for OCD intrusive doubt, which rises from 0.023 in the immediate-pre window to 0.072 in the immediate-post window. For bipolar disorder, Impulsive Risk and Crash reaches its highest value in the near-post window (0.143); for BPD, Self-Harm and Rage reaches its highest value in the later-post window (0.089); and for PTSD, Involuntary Reexperiencing reaches its highest value in the far-post window (0.062). These descriptive values indicate that, for several conditions, the most intense condition-specific difficulty language in the observed two-year span appears after the disclosure anchor rather than before it.

\begin{figure*}[t]
\centering
\includegraphics[width=0.8\textwidth]{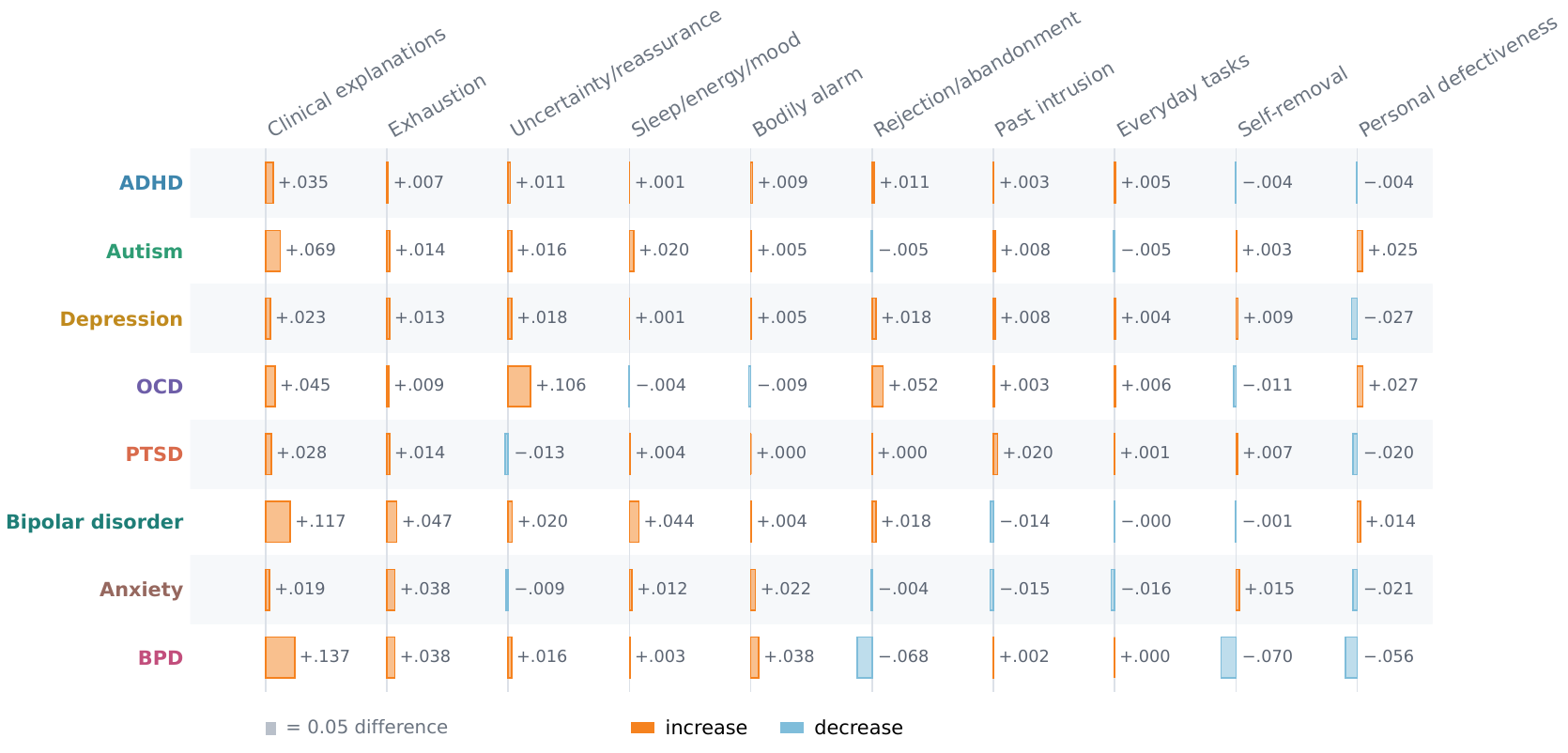}
\caption{Pooled late-pre-minus-early-pre differences for the ten primary-theme indicators. Column labels abbreviate the primary-theme names in Table~\ref{tab:theme-codebook}, ordered by mean difference across conditions.}
\Description{Grid of horizontal bars showing pooled late-pre-minus-early-pre differences for ten primary-theme indicators. The Seeking Clinical Explanations theme has the largest absolute difference in five condition profiles; the exhaustion theme does so for anxiety, the personal-defectiveness theme for depression, and the uncertainty theme for OCD.}
\label{fig:theme-change}
\end{figure*}

\findingclaim{The Seeking Clinical Explanations theme shows the largest early-to-late pre-disclosure difference in five of the eight conditions.}
While the specific nature of burden diverges across conditions, the posts written closest to disclosure share a common way of framing experience. For each primary-theme indicator, we subtract the pooled early-pre mean from the pooled late-pre mean. As shown by the prominent orange bars in Figure~\ref{fig:theme-change}, the \emph{Seeking Clinical Explanations} theme dominates this shift. It has the largest absolute pooled difference for ADHD (0.035), autism (0.069), PTSD (0.028), bipolar disorder (0.117), and BPD (0.137).

Among the other profiles, \emph{Feeling Too Exhausted for Daily Functioning} has the largest difference for anxiety (0.038), \emph{Interpreting Social Experiences as Evidence of Personal Defectiveness} has the largest absolute difference for depression ($-0.027$), and \emph{Trying to Eliminate Uncertainty Through Checking and Reassurance Seeking} has the largest difference for OCD (0.106). The corresponding pooled differences for \emph{Seeking Clinical Explanations} are 0.019 for anxiety, 0.023 for depression, and 0.045 for OCD. Figure~\ref{fig:theme-change} shows these pooled differences across all ten primary-theme indicators. The pattern also continues past the anchor: pooled across the immediate-post and near-post windows, the \emph{Seeking Clinical Explanations} indicator exceeds its late-pre mean for six of the eight conditions (all except OCD and BPD). While everyday difficulties remain condition-specific (e.g., masking for autism, reassurance-seeking for OCD), the most consistent cross-condition difference before the anchor is the larger share of posts that reach for clinical explanations in the late pre-disclosure windows. RQ3 next examines how these language indicators co-occur with recorded Reddit engagement.

\subsection{RQ3 (Community Reaction): Co-Occurrence of Recorded Engagement With Language Indicators}
\label{sec:results-rq3}
Having described the content of the surrounding posts, RQ3 examines how recorded engagement co-occurs with these language indicators. \findingclaim{Same-window correlations between language indicators and online community engagement on Reddit meet the corrected threshold only for ADHD and BPD.} The same-window analysis produces 360 Pearson correlations between the 15 language indicators and three non-duplicate engagement fields. Of these tests, 32 have raw $p \leq .05$, and 9 have $q \leq .05$ after Benjamini--Hochberg correction: 5 for ADHD and 4 for BPD, none for the other six conditions. Because authors contribute several author-windows each, these tests treat dependent observations as independent; BH correction controls the false discovery rate across the family but does not adjust for that dependence. We therefore read the retained associations alongside the author-level analysis reported below. The secondary Mann--Whitney family retains more corrected differences (75 of 360 tests at $q \leq .05$); because the zero threshold carries different meanings across indicators (Section~\ref{sec:methods}), we treat these distributional comparisons as context for the Pearson results rather than interpreting them individually. \emph{Presence share} is the proportion of author-windows in which a language-indicator value exceeds zero. One corrected association links the \emph{Seeking Clinical Explanations} indicator to upvote ratio for ADHD (presence share = 0.357, $r = .115$, $q < .001$). This sparsity indicates that, in these cohorts, higher values on the language indicators rarely correspond to consistent differences in recorded engagement.

\findingclaim{Higher language-indicator values co-occur with both higher and lower online community engagement on Reddit.}
Of the nine corrected correlations, 6 are positive and 3 are negative, spanning multiple language indicators and engagement fields. The retained set comprises affect indicators and primary-theme indicators. Notably, no association involving the burden dimension-share score meets the corrected threshold: within this design, the overall amount of condition-specific burden language shows no corrected association with any engagement field.

\begin{figure*}[t]
\centering
\includegraphics[width=0.67\textwidth]{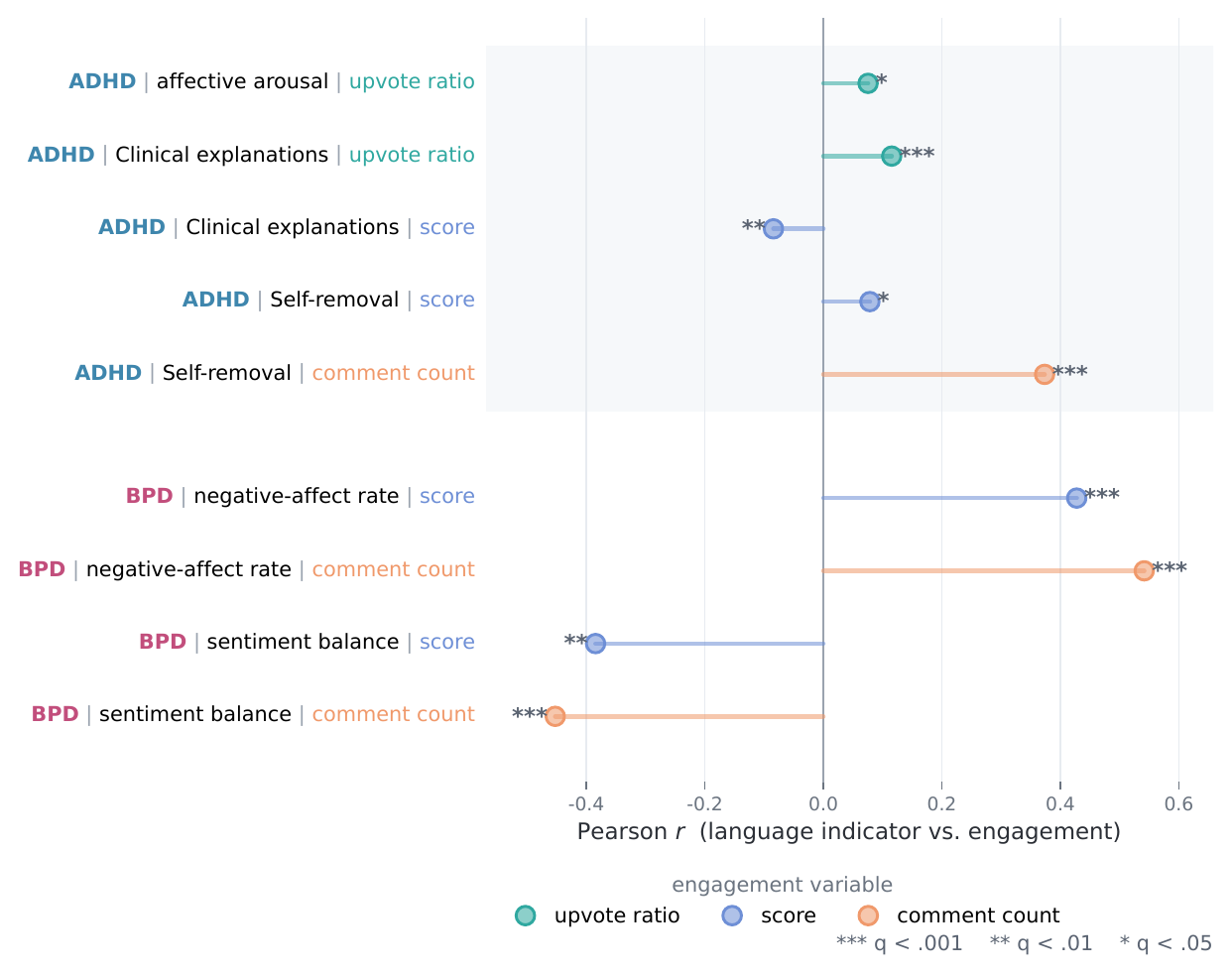}
\caption{The 9 of 360 same-window Pearson correlations with BH-adjusted \(q \leq .05\). Each row names the condition, the language indicator, and the engagement field; theme names are abbreviated as in Table~\ref{tab:theme-codebook}.}
\Description{Dotplot of same-window Pearson correlations that pass Benjamini--Hochberg q less than or equal to .05. Points right of zero indicate higher engagement as a language indicator increases, while points left of zero indicate lower engagement.}
\label{fig:engagement-dotplot}
\end{figure*}

Figure~\ref{fig:engagement-dotplot} visualizes this divergence. For ADHD, the \emph{Seeking Clinical Explanations} indicator is positively associated with upvote ratio and negatively with Reddit score; the \emph{Imagining Self-Removal as an Escape From Distress} indicator is positively associated with comment count and Reddit score; and affective arousal is positively associated with upvote ratio. For BPD, negative-affect rate is positively associated with comment count and Reddit score, while sentiment balance is negatively associated with both. As the opposing directions in Figure~\ref{fig:engagement-dotplot} show, higher language-indicator values do not uniformly co-occur with higher engagement; the corrected associations are instead specific to condition and engagement field, which cautions against treating visible engagement as a proxy for received support. Table~\ref{tab:2} in Appendix~\ref{app:author-aggregated-sensitivity} reports all 9 corrected associations with the author-windows behind each. These two cohorts survive correction for different reasons: ADHD contributes the most author-windows (2,063), so even small coefficients reach the corrected threshold (four of its five retained associations have $|r| < .12$), whereas BPD contributes the fewest (114) but shows the largest coefficients ($|r|$ = .38--.54). The remaining six conditions fall between these extremes (157--490 author-windows) with no coefficient exceeding $|r|$ = .17. What survives correction therefore tracks cohort size and effect magnitude rather than any clinical distinctiveness of ADHD or BPD. Two features of this retained set warrant caution: the self-removal theme exceeds zero in only 5.6\% of ADHD author-windows, so its correlations summarize a small active subset, and the BPD windows come from just 23 authors. Both should be read as descriptive associations within these subsets rather than stable condition-level effects.

\findingclaim{Author-level aggregation preserves correlation directions but leaves fewer correlations meeting the corrected threshold.}
After each author's observed windows are averaged into one row, all 9 corrected same-window associations retain the same direction, and 2 keep $q \leq .05$ after author-level BH adjustment. The retained associations are ADHD \emph{Imagining Self-Removal as an Escape From Distress} with comment count ($r = .265$, $q < .001$) and ADHD \emph{Seeking Clinical Explanations} with upvote ratio ($r = .181$, $q = .047$). Across the full 360-test author-level family, one additional corrected association does not appear in the corrected same-window set: BPD \emph{Experiencing the Body as Stuck in Alarm} with comment count ($r = .677$, $q = .047$). The other estimates no longer meet the corrected threshold. Appendix~\ref{app:author-aggregated-sensitivity} (Table~\ref{tab:3}) reports the full author-level sensitivity comparison for all nine corrected same-window associations.

\findingclaim{Only a small subset of correlations between one window's language indicators and next-window online community engagement on Reddit meets the corrected threshold.}
Across 216 lagged Pearson tests, 6 associations have $q \leq .05$. The \emph{Losing Control of Sleep, Energy, Mood, and Thought Speed} indicator is associated with next-window Reddit score for depression ($r = .253$, $q = .004$) and ADHD ($r = .111$, $q = .004$), and with next-window comment count for ADHD ($r = .097$, $q = .016$). For bipolar disorder, the \emph{Feeling Too Exhausted for Daily Functioning} indicator is associated with next-window Reddit score ($r = .319$, $q = .006$), and negative-affect rate is associated with next-window comment count ($r = .275$, $q = .027$). For ADHD, the \emph{Seeking Clinical Explanations} indicator is associated with next-window upvote ratio ($r = .091$, $q = .027$). These six coefficients are small to moderate, and the lagged analysis remains exploratory; language indicators in one window rarely show corrected associations with engagement in the next, and these associations do not establish prediction or causation.

\section{Discussion}
\label{sec:discussion}
The Disclosure Processes Model treats each self-disclosure as a single event within an ongoing process of sensemaking, rather than the static beginning of the experience being disclosed \citep{chaudoir2010disclosure}. Our findings translate this framing to Reddit; users produce condition-specific burden and explanation language months before formally disclosing a condition. Prior work on mental health often treats the disclosure event as the moment a condition becomes visible, using it to define ground-truth cohorts for predictive models \citep{chancellor2020methods, coppersmith2015adhd, kim2020deep} or to analyze the immediate social support that follows the announcement \citep{saha2020causal}. Our study instead unpacks how mental health experience becomes interpretable to others over time. Specifically, we map how \emph{language-visible burden} (RQ1), \emph{primary themes} (RQ2), and \emph{online community engagement on Reddit} (RQ3) evolve and intersect around a \emph{disclosure anchor}. The following sections introduce \emph{mental health self-disclosure visibility} to distinguish these forms of evidence and to draw out implications for research, online communities, and social media platforms like Reddit.

\subsection{The Visibility of Mental health Self-Disclosure}
\label{sec:self-disclosure-visibility}
\begin{figure*}[t]
   \centering
   \includegraphics[width=0.7\textwidth]{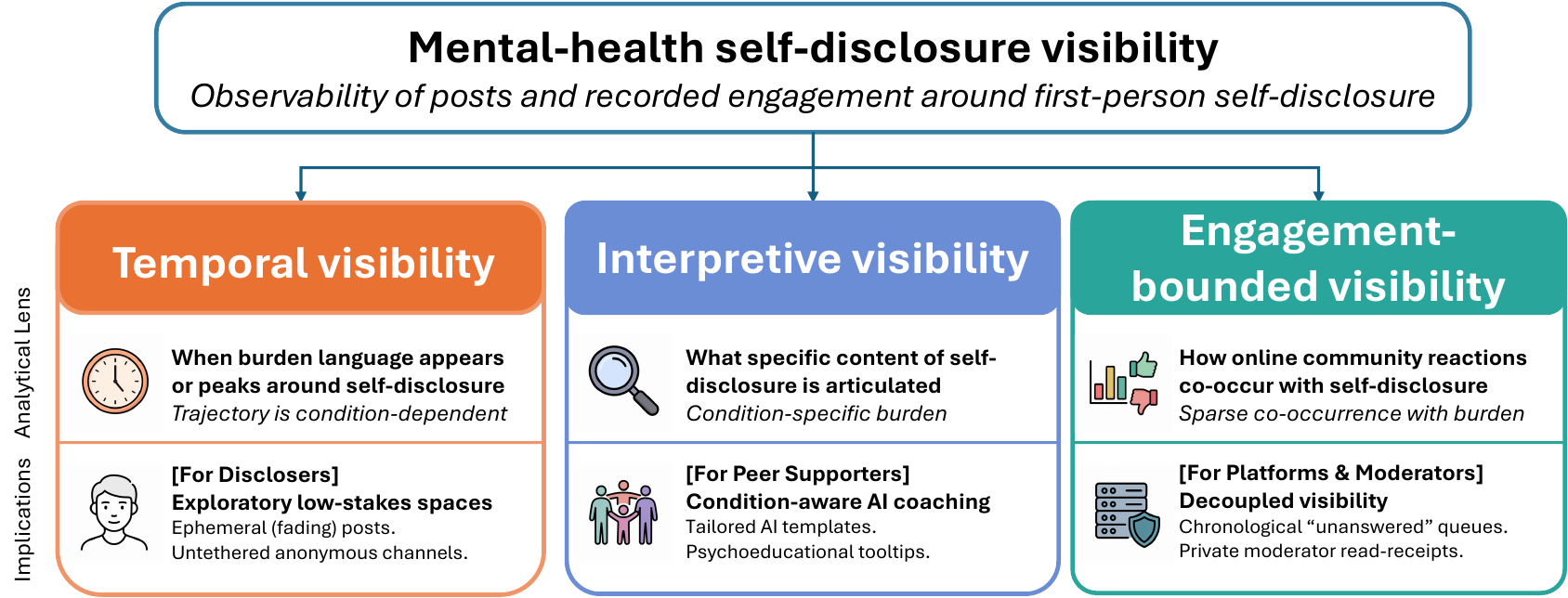}
   \caption{The mental health self-disclosure visibility framework: three dimensions of observability derived from RQ1--RQ3 (top) and their implications for online communities and platforms (bottom).}
   \Description{Concept diagram of the mental health self-disclosure visibility framework. A header box connects to three columns: temporal visibility, interpretive visibility, and engagement-bounded visibility. Each column contains an analytical-lens box summarizing what that dimension describes and an implications box listing the corresponding design and community implications for disclosers, peer supporters, and platforms or moderators.}
   \label{fig:self-disclosure-visibility}
\end{figure*}

Prior research on social media visibility shows that users distinguish the visibility of specific content from that of the person or identity behind it \citep{barta2024visibility}. Building on this distinction, we introduce \textit{mental health self-disclosure visibility} as a framework for understanding what researchers, algorithms, and online community members can observe in posts surrounding an explicit mental health disclosure. Language indicators capture what authors express in their posts but cannot establish their underlying experiences or intentions. Similarly, online community engagement on Reddit captures observable platform reactions but cannot establish whether community members responded with empathy or understanding. Figure~\ref{fig:self-disclosure-visibility} presents the framework’s three dimensions as an analytical lens (top row) and connects each dimension to implications for researchers, online communities, and platform designers (bottom row).

\textbf{Temporal Visibility.} First, temporal visibility positions explicit self-disclosure as a marker within an ongoing process rather than as the onset of a condition. We found that within the pre-disclosure year, the \emph{burden dimension-share score} peaked at different times across conditions—for example, during the immediate-pre period for ADHD but during the far-pre period for BPD. Across the full observation period, four conditions reached their highest condition-window mean only after disclosure. These heterogeneous trajectories complicate the linear assumption that diagnosis disclosure marks the point at which condition-related experiences first emerge—an assumption that can underlie computational models that align posting histories to detect condition onset \citep{de2016discovering, mansoor2024early} or treat diagnosis claims as fixed temporal anchors \citep{biester2022identity, ernala2019methodological}. Instead, consistent with the Disclosure Processes Model \citep{chaudoir2010disclosure}, users may express distress, uncertainty, or condition-related difficulties before explicitly naming a diagnosis, sometimes as a way of ``testing the waters'' \citep{andalibi2018testing, andalibi2016sexualabuse}. In our sample, condition-relevant burden language remained visible throughout the post-disclosure year.

\textbf{Interpretive Visibility.} Second, interpretive visibility highlights that the meanings conveyed around disclosure are condition-specific and socially situated. Our RQ2 findings show that the forms of burden expressed before disclosure varied across conditions, ranging from fear of abandonment among users disclosing BPD to reassurance-seeking among those disclosing OCD. Across five conditions, \emph{Seeking Clinical Explanations}---references to diagnoses, therapy, medication, and symptom labels---showed the largest pooled difference between the early- and late-pre-disclosure periods. Prior research on tensions surrounding online care \citep{eagle2023adhd, pendse2023marginalization}, neurodivergent identity performance \citep{leveille2024adhd}, and peer support \citep{sun2025ocd} suggests that diagnostic vocabulary does more than communicate clinical information: it can also help users interpret and legitimize their experiences and make those experiences intelligible to networked publics \citep{boyd2010networkedpublics, edwards2024autism}. Whereas earlier multi-condition detection research treated this language primarily as a set of predictive features \citep{cohan2018smhd, coppersmith2015adhd}, our findings are consistent with its broader role in sensemaking and identity negotiation.

\textbf{Engagement-Bounded Visibility.} Finally, engagement-bounded visibility distinguishes what users express from the observable reactions their posts receive, highlighting the limitations of relying on platform metrics of online community reaction alone. Our RQ3 findings showed sparse associations between language indicators and online community engagement on Reddit: only 9 of 360 same-window associations survived correction, none involved language-visible burden, and the observed patterns varied across conditions. Although prior work has linked audience engagement with the intimacy of subsequent disclosures \citep{ernala2018audience}, our findings caution against interpreting upvote ratios or comment counts as indicators of empathy or support quality. Community members’ responses are shaped by relational contexts \citep{andalibi2018responding}, and meaningful support is often conveyed through the content of responses rather than interaction counts alone \citep{andalibi2017sensitive}. Moreover, because users may have an “invisible audience” that views a post without visibly responding \citep{bernstein2013invisible}, limited engagement does not necessarily indicate an absence of attention or care. The interactions available to researchers are also shaped by algorithmic curation \citep{devito2017algorithms, milton2023tiktok} and the often-invisible labor of community moderation \citep{dosono2019moderation, matias2019civic}. Thus, recorded engagement represents only a partial and platform-mediated trace of how online communities encounter and respond to mental health disclosures.

\textit{Mental health self-disclosure visibility} shifts the research question from whether a mental health condition can be detected to what becomes visible and interpretable around an explicit self-disclosure. Our findings show that a common disclosure anchor does not make users’ surrounding experiences uniformly observable. Predictive models trained on disclosure-defined cohorts \citep{chancellor2020methods, coppersmith2015adhd} may therefore learn patterns of platform visibility—what users choose to express, when they express it, and how the platform and its communities respond—rather than the underlying clinical condition itself \citep{chancellor2023contextual, olteanu2019socialdata}. By distinguishing temporal patterns, thematic content, and community engagement, our framework preserves these differences rather than collapsing them into a single indicator of clinical status or severity. Self-disclosure data should therefore be understood not merely as an imperfect clinical label to be refined, but as a situated record of how users interpret, communicate, and negotiate mental health experiences within networked publics.

This lens also cautions against equating an absence of digital evidence with an absence of distress. Social media data reflect what users choose to express and what platform algorithms make visible, rather than the full range of users’ experiences \citep{barta2024visibility, tufekci2014bigquestions}. Peer-support systems \citep{sharma2023human, song2025typing} and algorithmic moderation tools \citep{kumar2024watch, steiger2021psychological} that prioritize highly visible or highly engaged disclosures may therefore overlook users who express distress indirectly or receive little community response, particularly before explicit disclosure.

\subsection{Implications for Online Community Practice and Social Media Platform Design}
\label{sec:implications}
Our findings describe what becomes visible in social media users’ language around first-person mental health disclosures and how recorded engagement corresponds with that language. These patterns have distinct implications for three stakeholders in this networked process: people who disclose, community members who respond, and platforms and moderators, who shape how content circulates and receives attention. Figure~\ref{fig:self-disclosure-visibility} summarizes these stakeholder-specific implications.

\textbf{People who disclose: Supporting an extended sensemaking process.} Our findings position a disclosure post within a longer trajectory of interpreting and communicating mental health experiences. Across all eight mental health conditions, condition-relevant burden language was measurable in authors’ posts months before the selected mental health disclosure. These patterns are consistent with disclosure-process research that situates disclosure within an extended process of appraisal and communication \citep{chaudoir2010disclosure}. They also extend event-aligned research on depression, which found that anxiety and sadness language increased before diagnosis claims \citep{biester2022identity}, by demonstrating heterogeneous trajectories across multiple conditions: within the pre-disclosure year, language-visible burden peaked immediately before disclosure for six mental health conditions, earlier for PTSD, and furthest from disclosure for BPD.

Community resources can recognize posts preceding explicit disclosure as part of an ongoing sensemaking process. Prior qualitative research in ADHD and autism communities describes disclosure as identity work that may involve comparing one’s experiences with those of others \citep{eagle2023adhd,leveille2024adhd,edwards2024autism}. \textbf{Design Implications: }Community wikis, pinned guides, and peer-authored FAQs could acknowledge that uncertainty, exploratory questions, and indirect descriptions of difficulty commonly precede explicit self-disclosure. Such resources could help community members respond constructively to posts such as “Does anyone else experience this?” without pressuring authors to adopt a diagnostic label. Similarly, systems that require diagnostic tags or flairs to access relevant resources may inadvertently create pressure to claim a label before a user is ready. Low-stakes affordances, such as optional anonymous channels or ephemeral posts, could provide spaces for users to “test the waters” while retaining greater control over whether a mental health label becomes attached to their persistent digital identity \citep{andalibi2018testing,ma2016anonymity}.

The sparse correspondence between language indicators and recorded engagement also suggests that people who disclose should not be encouraged to interpret visible reactions as a direct measure of whether their experiences matter or have been understood. Low engagement provides limited information about a post’s audience because users may underestimate how many people encounter their content \citep{bernstein2013invisible}, and platform mechanisms further obscure who sees or responds to a post \citep{barta2024visibility}. Prior work has identified potential benefits of disclosure, including therapeutic value and reduced self-stigma \citep{corrigan2013reducing,ernala2017linguistic}, but these benefits should not be framed as contingent on receiving a large number of comments or upvotes. Community education and design can help users distinguish the personal decision to disclose from expectations about visible platform engagement.

\textbf{Peer supporters and community members: Responding to varied expressions of burden.} Peer support is a central function of online mental health communities \citep{naslund2016future,dechoudhury2014reddit,o2018suddenly}. The leading pre-disclosure burden dimensions differ across cohorts: masking and sensory overload for autism, reassurance-seeking for OCD, Abandonment Fear for BPD, and forgetfulness and disorganization for ADHD. These differences suggest that effective peer support cannot rely on a single generic model of distress. Instead, supporters can respond to the specific experiences described in a post without assuming that those expressions establish a diagnosis. This approach complements evidence that effective support involves adapting language to the support-seeker \citep{sharma2018mental} and that psychosocial outcomes depend on the content of responses, not simply their number \citep{saha2020causal}. \textbf{Design Implications: }Peer-support designs could provide community-specific guidance while preserving users’ autonomy and avoiding diagnostic inference. Human--AI coaching systems and automated response prompts \citep{sharma2023human,song2025typing}, for example, could offer optional guidance based on the community context or concerns explicitly described by the user. In an OCD-focused community, a responder might voluntarily access community- and clinician-reviewed guidance on validating distress without reinforcing cycles of reassurance-seeking \citep{sun2025ocd}. Such tools should support responders’ judgment rather than automatically classify authors or presume an undisclosed condition.

Peer-support practices should also account for the possibility that posts expressing gradual or less affectively intense difficulties receive fewer visible responses than crisis-adjacent posts. Some of the surviving engagement associations involved affectively intense language, but these patterns were concentrated in a small number of conditions and were not uniformly robust across analytic checks. In contrast, language-visible burden itself showed no association that met the corrected threshold. These findings do not establish that quieter expressions are systematically ignored, particularly given the cohort sizes and unmeasured differences in exposure. They nevertheless identify a plausible risk: if community responses concentrate around posts that have already attracted attention, expressions of slower-building burden may remain unanswered. Structured check-in threads, norms encouraging replies to zero-comment posts, and dedicated peer-support formats could broaden opportunities for response without requiring a post to first become popular.

\textbf{Platform moderators, designers, and public-health practice: Engagement is not a triage signal.} Moderators sustain mental health communities, often at substantial emotional cost \citep{dosono2019moderation,matias2019civic,steiger2021psychological}. Our cohort sizes and lack of adjustment for exposure prevent us from concluding that engagement and burden are unrelated, but the condition-specific association findings give platforms no basis for assuming that highly engaged posts reflect greater burden or that less-engaged posts reflect less need.

\textbf{Design Implications: }Because popularity-based ranking may make posts with little initial engagement more difficult to encounter \citep{barta2024visibility,chancellor2023contextual}, platforms could create chronological “unanswered” queues, structured check-in spaces, or optional moderator acknowledgments. Language indicators could also be explored for aggregate community monitoring, but they require ongoing validation as community language changes \citep{chancellor2016thyghgapp}. They should not be repurposed to identify individuals as “pre-disclosure.” These indicators are not clinical measures \citep{chancellor2020methods,ernala2019methodological}, and inferring undisclosed diagnoses could undermine disclosure autonomy \citep{chaudoir2010disclosure} and contribute to algorithmic stigmatization \citep{andalibi2023conceptualizing}.

More broadly, disclosure language and Reddit engagement fields should not independently determine clinical need, risk, or priority for outreach. When such signals inform consequential moderation or support decisions, platforms should communicate their limitations, incorporate contextual and human review, and evaluate systems across conditions and communities. Rather than intensifying surveillance of “quiet” users, platforms can provide voluntary, low-friction access to credible information, peer support, and crisis resources without requiring diagnostic language or high engagement. \textit{Mental health self-disclosure visibility} thus clarifies where digital evidence ends, helping stakeholders use observable traces without treating experiences that remain invisible in the data as absent from users’ lives.

\section{Limitations, Ethics, and Future Work}
\label{sec:limitations-ethics-future-work}
Five limitations bound what these findings support. First, disclosure-anchor selection may miss earlier disclosures absent from the retrieved post history, does not reassign authors whose posts meet another condition's criteria, and prioritizes precision over cohort size through the exclusions in Section~\ref{sec:methods}. The selected anchor is the earliest eligible same-condition post at or before the initial candidate, not the author's first-ever disclosure or diagnosis date, and because only the anchor is removed, other disclosure posts may remain in the timeline; the comparisons therefore describe language relative to the selected event. Second, the amount of available pre-disclosure history varies across authors: 55 retained authors have fewer than five pre-disclosure posts (19 have none) and contribute only to the event-time windows in which they have observed posts. Third, the lexical indicators may miss context, sarcasm, negation, and relevant language outside the predefined pattern sets, and they do not measure clinical severity or diagnostic thresholds. Human validation further showed that coder agreement was substantial rather than near-perfect, and fewer than half of the automatically identified primary-theme instances were confirmed, so the primary-theme indicators support weaker claims than the burden dimensions. The fixed validation samples also assess lexical assignments rather than anchor selection or prevalence, and primary-theme indicators do not establish authors’ intentions. Our computational checks verify the declared rules and equations rather than semantic validity. 

Fourth, condition-window means and pooled early- and late-pre-disclosure estimates aggregate across authors, so no reported maximum or difference establishes within-author change, symptom onset, or an effect caused by disclosure. Fifth, Reddit users are not representative of all people with mental health conditions. The engagement fields are unadjusted for subreddit, calendar time, post exposure, author activity, and platform changes, so same-window correlations describe co-occurrences between specific language indicators and recorded engagement fields rather than a general community-support response. The same-window tests also treat author-windows as independent, so their $q$-values do not account for within-author dependence; the author-level analysis is the corresponding check. The lagged analyses cannot rule out confounding and do not establish prediction or causality. This analysis uses public Reddit posts about sensitive mental health topics. The posts were already publicly available when we collected them, and we had no interaction with the people who wrote them; our institution suggests that institutional review board review is not required for secondary analysis of public data of this kind. Two extensions follow from these limits. Qualitative analysis of subtheme variation, boundary cases, and narrative depth would address what automated primary-theme indicators cannot resolve. Analyzing comment content, rather than engagement counts, would establish whether community responses constitute support, normalization, or harm.

\section{Conclusion}
\label{sec:conclusion}
This study examined what becomes visible in Reddit users' language and in recorded community engagement around a first-person mental health diagnosis disclosure. Across eight mental health conditions, the timing and content of language-visible burden are condition-specific: within the pre-disclosure year, most conditions peak in the month before disclosure, and burden language remains visible across the post-disclosure year. The \emph{Seeking Clinical Explanations} theme shows the most consistent difference between the early and late pre-disclosure periods, whereas corrected associations between language indicators and Reddit engagement fields are sparse and limited to two conditions. Together, these findings support \emph{mental health self-disclosure visibility} as a way to study disclosure as an unevenly visible process, and the event-time design behind it keeps author language and recorded engagement analytically distinct for digital mental health, social computing, HCI, and computational social science research.

\bibliographystyle{ACM-Reference-Format}
\bibliography{references}

\appendix

\section{Search Keywords}
\label{app:search-keywords}

Table~\ref{tab:search-terms} lists the query used in Step 1 of Section~\ref{sec:data-collection}. A post entered the candidate set only if it contained at least one of the 41 first-person disclosure phrases and at least one of the 33 mental health terms; terms combined disjunctively within each list and conjunctively across the two. The phrases fall into two families because first-person diagnosis reports take two grammatical forms, one placing the author in the subject position (``I was diagnosed with'') and one placing a clinician there (``my psychiatrist diagnosed me with''). The term list is broader than the eight conditions analyzed here because we did not fix those conditions before collection; Section~\ref{sec:data-collection} gives the criterion by which the eight were retained. Coverage is uneven in one respect. Seven conditions have condition-specific terms, marked in boldface, but BPD does not, entering only through the general terms or through ``personality disorder;'' BPD is also the smallest cohort, contributing 23 authors to the corrected associations in Section~\ref{sec:results-rq3}. This query is a recall-oriented first pass rather than a definition of disclosure, and it cannot separate an author's own diagnosis from one that is quoted, hypothetical, or attributed to someone else. The criteria that draw that distinction---an explicit first-person self-attribution, a single named study condition, and the absence of uncertainty wording---are applied afterward in Section~\ref{sec:cohort-construction} and Appendix~\ref{app:disclosure-anchor-criteria}, reducing the 83,140 authors this query returned to the 739 analyzed here.

\begin{table*}[t]
\centering
\footnotesize
\caption{Search terms for identifying candidate self-disclosure posts.}
\label{tab:search-terms}
\begin{tabular}{@{}p{0.27\textwidth}p{0.65\textwidth}@{}}
\toprule
\multicolumn{2}{@{}l}{\textbf{Self-disclosure terms (41 phrases)}} \\
\midrule
Diagnosis statements (19) & ``was / got / am / have been diagnosed with,'' ``officially / clinically / formally / previously diagnosed with,'' ``received a diagnosis of,'' ``received diagnosis,'' ``was given a diagnosis of,'' ``have / has / had a diagnosis of,'' ``has an official / clinical diagnosis of,'' ``carry a diagnosis of,'' ``was clinically / medically determined to have'' \\
Clinician-referenced statements (22) & ``diagnosed me with,'' ``diagnosed by a psychiatrist / psychologist / doctor / therapist,'' ``my doctor / psychiatrist / psychologist / therapist diagnosed me with,'' ``my doctor / psychiatrist / psychologist said / says I have / had,'' ``my therapist told me'' \\
\midrule
\multicolumn{2}{@{}l}{\textbf{Mental health terms (33 terms)}} \\
\midrule
General terms & mental health; mental disorder; mental illness \\
Study-condition terms & \textbf{ADHD}; \textbf{attention deficit}; \textbf{attention-deficit hyperactivity disorder}; \textbf{anxiety disorder}; \textbf{autism spectrum disorder}; \textbf{asperger's syndrome}; \textbf{autistic}; \textbf{bipolar disorder}; \textbf{manic depression}; \textbf{cyclothymia}; \textbf{depression}; \textbf{depressive disorder}; \textbf{MDD}; \textbf{OCD}; \textbf{obsessive-compulsive disorder}; \textbf{PTSD}; \textbf{post-traumatic stress disorder} \\
Other queried terms & schizophrenia; paranoid schizophrenia; schizo; panic disorder; social phobia; eating disorder; anorexia; bulimia; personality disorder; idiopathic developmental intellectual disability; conduct disorder; anger issues; aggressive behavior \\
\bottomrule
\end{tabular}
\smallskip

\parbox{0.92\textwidth}{\footnotesize\emph{Note.} Posts required at least one term from each section. Boldface marks the eight study conditions; BPD was captured through general and personality-disorder terms. Parentheses give case-insensitive pattern counts, and slashes denote alternatives expanded as separate patterns.}
\end{table*}

\section{Author-Aggregated Sensitivity Results}
\label{app:author-aggregated-sensitivity}

This appendix reports selected author-level sensitivity results summarized in Section~\ref{sec:results}. The analysis changes the unit from author-window to author by averaging each author's observed event-time windows into one row before re-estimating Pearson correlations and BH-adjusted $q$-values. The table shows that coefficient signs stay consistent across the two units and that most, though not all, coefficients shrink once repeated windows within authors are removed.

\begin{table*}[t]
\centering
\footnotesize
\setlength{\tabcolsep}{1.0pt}
\renewcommand{\arraystretch}{1.03}
\caption[Same-Window Correlations Meeting the BH-Adjusted Threshold]{Same-Window Correlations Meeting the BH-Adjusted Threshold. \emph{Note.} The table reports the 9 of 360 same-window Pearson correlations with BH-adjusted \(q \leq .05\); all 15 language indicators were tested. \texttt{Presence share} is defined in Section~\ref{sec:methods}. $n$ is the number of author-windows contributing to each correlation.}
\label{tab:2}
\begin{adjustbox}{max width=\textwidth}
\begin{tabular}{@{}>{\raggedright\arraybackslash}p{0.09\textwidth}
>{\raggedright\arraybackslash}p{0.46\textwidth}
>{\raggedright\arraybackslash}p{0.12\textwidth}
>{\raggedright\arraybackslash}p{0.085\textwidth}
>{\raggedright\arraybackslash}p{0.05\textwidth}
>{\raggedright\arraybackslash}p{0.07\textwidth}
>{\raggedright\arraybackslash}p{0.06\textwidth}@{}}
\toprule
\textbf{Condition} & \textbf{Language indicator} & \textbf{Engagement field} & \textbf{Presence share} & \textbf{$r$} & \textbf{BH $q$} & \textbf{$n$} \\
\midrule
ADHD & \emph{Imagining Self-Removal as an Escape From Distress} & comment count & 0.056 & .373 & $< .001$ & 2,063 \\
ADHD & \emph{Seeking Clinical Explanations} & upvote ratio & 0.357 & .115 & $< .001$ & 2,063 \\
ADHD & \emph{Seeking Clinical Explanations} & Reddit score & 0.357 & $-.084$ & $.007$ & 2,063 \\
ADHD & \emph{Imagining Self-Removal as an Escape From Distress} & Reddit score & 0.056 & .079 & $.016$ & 2,063 \\
ADHD & affective arousal & upvote ratio & 0.292 & .075 & $.024$ & 2,063 \\
BPD & negative-affect rate & comment count & 0.553 & .541 & $< .001$ & 114 \\
BPD & sentiment balance & comment count & 0.237 & $-.452$ & $< .001$ & 114 \\
BPD & negative-affect rate & Reddit score & 0.553 & .428 & $< .001$ & 114 \\
BPD & sentiment balance & Reddit score & 0.237 & $-.384$ & $.001$ & 114 \\
\bottomrule
\end{tabular}
\end{adjustbox}
\end{table*}

\begin{table*}[t]
\centering
\scriptsize
\setlength{\tabcolsep}{1.0pt}
\renewcommand{\arraystretch}{1.03}
\caption[Selected Author-Level Sensitivity Results]{Selected Author-Level Sensitivity Results. \emph{Note.} Author-window columns reproduce the nine corrected same-window correlations (Table~\ref{tab:2}); author-level columns re-estimate them after collapsing each author's observed windows to one mean row.}
\label{tab:3}
\begin{adjustbox}{max width=\textwidth}
\begin{tabular}{@{}>{\raggedright\arraybackslash}p{0.080\textwidth}
>{\raggedright\arraybackslash}p{0.35\textwidth}
>{\raggedright\arraybackslash}p{0.090\textwidth}
>{\raggedright\arraybackslash}p{0.075\textwidth}
>{\raggedright\arraybackslash}p{0.085\textwidth}
>{\raggedright\arraybackslash}p{0.075\textwidth}
>{\raggedright\arraybackslash}p{0.085\textwidth}
>{\raggedright\arraybackslash}p{0.055\textwidth}
>{\raggedright\arraybackslash}p{0.055\textwidth}@{}}
\toprule
\textbf{Condition} & \textbf{Language indicator} & \textbf{Engagement field} & \textbf{Window $r$} & \textbf{Window BH $q$} & \textbf{Author $r$} & \textbf{Author BH $q$} & \textbf{Same sign} & \textbf{Author BH $q \leq .05$} \\
\midrule
ADHD & \emph{Imagining Self-Removal as an Escape From Distress} & comment count & .373 & $< .001$ & .265 & $< .001$ & yes & yes \\
ADHD & \emph{Seeking Clinical Explanations} & upvote ratio & .115 & $< .001$ & .181 & $.047$ & yes & yes \\
BPD & negative-affect rate & comment count & .541 & $< .001$ & .412 & $.808$ & yes & no \\
BPD & sentiment balance & comment count & $-.452$ & $< .001$ & $-.329$ & $.810$ & yes & no \\
BPD & negative-affect rate & Reddit score & .428 & $< .001$ & .179 & $.850$ & yes & no \\
BPD & sentiment balance & Reddit score & $-.384$ & $.001$ & $-.285$ & $.810$ & yes & no \\
ADHD & \emph{Seeking Clinical Explanations} & Reddit score & $-.084$ & $.007$ & $-.147$ & $.240$ & yes & no \\
ADHD & \emph{Imagining Self-Removal as an Escape From Distress} & Reddit score & .079 & $.016$ & .006 & $.991$ & yes & no \\
ADHD & affective arousal & upvote ratio & .075 & $.024$ & .167 & $.094$ & yes & no \\
\bottomrule
\end{tabular}
\end{adjustbox}
\end{table*}

\section{Computational Measurement Details}
\label{app:measurement}

The main text summarizes the study design, measurement rationale, and statistical families. This appendix reports the equations implemented by the deterministic pipeline. Consistent with computational measurement guidance, each equation defines an observable indicator and its inferential boundary \citep{grimmer2013text,jacobs2021measurement}. The burden measures are language indicators rather than clinical measures, and the Reddit engagement fields do not measure exposure or support. Table~\ref{tab:appendix-measurement-map} maps the reported outputs to their operational equations and grounding references.

\begin{table*}[t]
\centering
\footnotesize
\setlength{\tabcolsep}{2.0pt}
\renewcommand{\arraystretch}{1.12}
\setlength{\emergencystretch}{2em}

\caption[Appendix Measurement-to-Output Reference Map]{Appendix Measurement-to-Output Reference Map. \emph{Note.} This table links each manuscript table or visualization to the appendix equations and citations used to define or calculate the relevant measurements.}
\label{tab:appendix-measurement-map}

\begin{tabular}{@{}>{\raggedright\arraybackslash}p{0.165\textwidth}
>{\raggedright\arraybackslash}p{0.220\textwidth}
>{\raggedright\arraybackslash}p{0.3\textwidth}
>{\raggedright\arraybackslash}p{0.27\textwidth}@{}}
\toprule
\textbf{Output reference} & \textbf{Measurement object} & \textbf{Equation reference codes} & \textbf{Grounding references} \\
\midrule
Table~\ref{tab:1}; Figure~\ref{fig:peak-timing} & Condition profile and burden dimension-share score & Event-time indexing in Eq.~\eqref{eq:event-time}; burden-dimension hit and dimension-share score in Eqs.~\eqref{eq:burden-dimension-hit}--\eqref{eq:burden-dimension-share}; author-window aggregation and intervals in Eqs.~\eqref{eq:author-window-mean}--\eqref{eq:author-window-ci} & Longitudinal event-time analysis \citep{singer2003longitudinal,dechoudhury2013postpartum}; proxy measurement and construct validity \citep{grimmer2013text,jacobs2021measurement,messick1995validity} \\

Figure~\ref{fig:burden-heatmap} & Burden-dimension hit-rate heatmap & Burden-dimension hits in Eq.~\eqref{eq:burden-dimension-hit}; author-window aggregation in Eq.~\eqref{eq:author-window-mean} & Clinical-domain grounding for the study-specific dimensions \citep{apa2022dsm,baroncohen2001aq,blevins2015pcl5,bohus2009bsl23,goodman1989ybocs,hirschfeld2000mdq,hull2017camouflaging,kessler2005asrs,kroenke2001phq9,spitzer2006gad7} \\

Figure~\ref{fig:theme-change} & Primary-theme indicator difference & Primary-theme hit in Eq.~\eqref{eq:primary-theme-hit}; author-window aggregation in Eq.~\eqref{eq:author-window-mean}; pooled mean in Eq.~\eqref{eq:pooled-mean} & Thematic analysis and computational grounded theory \citep{braun2006thematic,nelson2020cgt} \\

Table~\ref{tab:2}; Figure~\ref{fig:engagement-dotplot} & Same-window correlations with Reddit engagement fields & Pearson correlation in Eq.~\eqref{eq:same-window-correlation}; presence share in Eq.~\eqref{eq:indicator-zero-split}; burden dimension-share score, affect indicators, and primary-theme indicators in Eqs.~\eqref{eq:burden-dimension-share}, \eqref{eq:affect-rate}--\eqref{eq:affective-arousal}, and \eqref{eq:primary-theme-hit}; BH adjustment in Eq.~\eqref{eq:bh-q-value} & Research on Reddit response and support quality \citep{andalibi2017sensitive,dechoudhury2014reddit}; Benjamini--Hochberg adjustment \citep{benjamini1995fdr}; affect lexicon traditions \citep{hutto2014vader,mohammad2013nrc,pennebaker2015liwc} \\

Table~\ref{tab:3} & Author-level sensitivity correlations & Author-level correlation in Eq.~\eqref{eq:author-level-correlation}; BH adjustment in Eq.~\eqref{eq:bh-q-value} & Observational social-media data research \citep{olteanu2019socialdata,ruths2014social}; Benjamini--Hochberg adjustment \citep{benjamini1995fdr} \\

Exploratory lagged check in Section~\ref{sec:results} & Next-window correlation with Reddit engagement fields & Lagged correlation in Eq.~\eqref{eq:lagged-correlation}; BH adjustment in Eq.~\eqref{eq:bh-q-value} & Longitudinal contrast without causal interpretation \citep{singer2003longitudinal}; observational social-media data research \citep{olteanu2019socialdata,ruths2014social} \\
\bottomrule
\end{tabular}
\end{table*}

\subsection{Event-Time Indexing}
\label{app:event-time-indexing}

For each surrounding post \emph{i} by author \emph{a}, event time is the number of days between the post timestamp and the disclosure-anchor timestamp:

\begin{equation*}
d_i = \frac{\mathrm{timestamp}_i - \mathrm{disclosureTimestamp}_a}{86,400}
\tag{A1}
\label{eq:event-time}
\end{equation*}

The denominator converts seconds to days. Event-time windows are then assigned from this value using the fixed intervals described in Section~\ref{sec:methods}. Each event-time interval includes the lower but not the upper boundary.

\subsection{Operational Criteria for Disclosure-Anchor Selection}
\label{app:disclosure-anchor-criteria}

The selection procedure separates an audit score from the binary eligibility criteria. For each post $i$ and assigned condition $c$, the score records condition mentions in the title and body that pass the context rules, a condition-related subreddit cue, and a qualifying first-person statement linking the condition to the author. The score makes the recorded evidence inspectable, but it does not determine eligibility, and no minimum score threshold is applied \citep{grimmer2013text,ratner2017snorkel}.

\begin{equation*}
F_{i,c} = 8D_{i,c}
\tag{A2}
\label{eq:disclosure-evidence-score}
\end{equation*}

\begin{equation*}
S_{i,c} = 4\min(T_{i,c},2) + \min(Q_{i,c},3) + R_{i,c} + F_{i,c}
\tag{A3}
\label{eq:condition-match-score}
\end{equation*}

Here, $T_{i,c}$ and $Q_{i,c}$ count named-condition mentions in the title and body that pass the context rules, respectively; $R_{i,c}$ equals 3 when a condition-related subreddit cue is present and 0 otherwise; and $D_{i,c}$ indicates at least one qualifying first-person statement linking the condition to the author. We retain $S_{i,c}$ as an audit field and use it only to order eligible candidates with the same timestamp, not as an independent basis for eligibility.

Let $K_{i,c}$ indicate that $c$ is the only named study condition in the post. The initial binary gate requires both a qualifying first-person statement and this single-condition criterion:

\begin{equation*}
B_{i,c} =
\mathbf{1}\left[D_{i,c}=1 \land K_{i,c}=1\right]
\tag{A4}
\label{eq:candidate-disclosure-gate}
\end{equation*}

The statement represented by $D_{i,c}$ cannot express uncertainty; deny, question, or rule out the condition; quote or relay another person's disclosure; or attribute the condition to another person. Subjectless diagnosis wording and symptom-only wording also yield $D_{i,c}=0$. Let $U_{i,c}$ indicate uncertainty about the assigned condition anywhere in the post. An eligible disclosure post must pass the binary gate and contain no such wording:

\begin{equation*}
C^{\mathrm{eligible}}_{i,c} =
\mathbf{1}\left[
B_{i,c}=1
\land U_{i,c}=0
\right]
\tag{A5}
\label{eq:eligible-disclosure-label}
\end{equation*}

Before selecting an anchor, we exclude an author when the initial candidate names multiple study conditions or expresses uncertainty about the assigned condition. For each remaining author, let $c_a$ denote author $a$'s assigned condition, $t_a^0$ the timestamp of the initial candidate disclosure, and $P_a$ the posts available for that author after study preprocessing, including the initial candidate disclosure. The selected disclosure anchor is the earliest eligible post for $c_a$ with a timestamp no later than $t_a^0$:

\begin{equation*}
i_a^* =
\underset{i\in P_a:\,t_i\leq t_a^0,\;C^{\mathrm{eligible}}_{i,c_a}=1}
{\operatorname{arg\,min}}\;t_i
\tag{A6}
\label{eq:disclosure-anchor-selection}
\end{equation*}

If this set is empty, the author is excluded from the final cohort. When eligible posts share the earliest timestamp, the condition-match score orders the tied candidates before the post identifier breaks any remaining tie. The rule holds $c_a$ fixed: a post naming another study condition does not reassign the author and cannot serve as the selected anchor. Thus, $i_a^*$ is the earliest eligible same-condition post identified within the available study history and the stated temporal boundary, not a verified diagnosis date or first-ever disclosure \citep{chancellor2020methods,guntuku2017detecting}.

\subsection{Condition-Specific Burden Dimensions}
\label{app:burden-dimensions}

Table~\ref{tab:burden-dimensions} lists the burden dimensions defined for each condition, their definitions, and the clinical literature that informed them.

\begin{table*}[t]
\centering
\scriptsize
\setlength{\tabcolsep}{3pt}
\renewcommand{\arraystretch}{1.08}
\setlength{\emergencystretch}{2em}

\caption[Condition-Specific Burden Dimensions]{Condition-Specific Burden Dimensions. \emph{Note.} Definitions describe the post language each dimension is intended to capture. Literature informed the domains, but this study's dimensions and search terms do not reproduce instrument subscales or scores.}
\label{tab:burden-dimensions}

\begin{tabular}{@{}
>{\raggedright\arraybackslash}p{0.07\textwidth}
>{\raggedright\arraybackslash}p{0.62\textwidth}
>{\raggedright\arraybackslash}p{0.28\textwidth}@{}}
\toprule
\textbf{Condition} & \textbf{Dimensions and definitions} & \textbf{Sources} \\
\midrule
ADHD &
\textbf{Inattention:} difficulty sustaining attention or keeping the mind from wandering.\newline
\textbf{Task Follow-Through:} difficulty starting or finishing tasks.\newline
\textbf{Organizational Lapses:} forgetting, losing track, missing deadlines, or difficulty organizing.\newline
\textbf{Hyperactive Impulsivity:} fidgeting, difficulty sitting still, impulsivity, or interrupting. &
Adult ADHD Self-Report Scale (ASRS) items \citep{kessler2005asrs}. \\

Autism &
\textbf{Social Opacity:} difficulty interpreting social cues, rules, or intentions.\newline
\textbf{Social Camouflaging:} hiding autistic traits or consciously compensating in social settings.\newline
\textbf{Sensory Overload:} sensory sensitivity, overstimulation, meltdown, or shutdown.\newline
\textbf{Change Intolerance:} reliance on predictability or difficulty with transitions and change. &
Autism-Spectrum Quotient (AQ) social and attention-switching content, the \emph{Diagnostic and Statistical Manual of Mental Disorders}, Fifth Edition, Text Revision (DSM-5-TR), and camouflaging research \citep{apa2022dsm,baroncohen2001aq,hull2017camouflaging}. \\

Depression &
\textbf{Low Mood:} sadness or sustained low-mood language.\newline
\textbf{Anhedonia:} reduced interest or pleasure, including emotional numbness when it conveys diminished pleasure.\newline
\textbf{Functional Fatigue:} low energy, exhaustion, or difficulty with everyday activities.\newline
\textbf{Self-Devaluation:} self-devaluation, failure, guilt, shame, or describing oneself as a burden.\newline
\textbf{Self-Harm Ideation:} self-harm, suicidal thinking, or not wanting to live. &
Patient Health Questionnaire-9 (PHQ-9) items and DSM-5-TR depressive descriptions \citep{apa2022dsm,kroenke2001phq9}. \\

OCD &
\textbf{Intrusive Doubt:} unwanted obsessive thoughts, persistent doubt, or difficulty stopping a concern.\newline
\textbf{Compulsive Rituals:} repeated checking, compulsions, or acts intended to reduce discomfort.\newline
\textbf{Reassurance Seeking:} seeking reassurance or complete confirmation.\newline
\textbf{Distressing Interference:} language that describes obsessive-compulsive experiences as distressing or disruptive. &
Yale--Brown Obsessive Compulsive Scale (Y-BOCS) obsession, compulsion, interference, and distress structure, interpreted with DSM-5-TR OCD descriptions \citep{apa2022dsm,goodman1989ybocs}. \\

PTSD &
\textbf{Involuntary Reexperiencing:} flashbacks, nightmares, triggers, or involuntary reexperiencing.\newline
\textbf{Avoidant Shutdown:} avoiding reminders, shutting down, dissociating, or emotional numbing.\newline
\textbf{Negative Trauma Cognitions:} trauma-related guilt, shame, blame, or persistent beliefs that the world is unsafe.\newline
\textbf{Hyperarousal:} hypervigilance, exaggerated startle, sleep disruption, or angry outbursts. &
PTSD Checklist for DSM-5 (PCL-5) items and DSM-5-TR PTSD domains \citep{apa2022dsm,blevins2015pcl5}. \\

Bipolar disorder &
\textbf{Decreased Sleep Need:} reduced need for sleep or remaining awake without feeling tired, rather than generic insomnia alone.\newline
\textbf{Accelerated Thinking:} accelerated thinking, rapid speech, or unusually many ideas.\newline
\textbf{Manic Activation:} manic or hypomanic activation, euphoria, or unusually high energy.\newline
\textbf{Impulsive Risk and Crash:} risky or impulsive action during activation, or an abrupt drop in mood or energy after a period of activation. &
Mood Disorder Questionnaire (MDQ) activation items and DSM-5-TR bipolar descriptions \citep{apa2022dsm,hirschfeld2000mdq}. \\

Anxiety &
\textbf{Uncontrollable Worry:} persistent worry or difficulty stopping rumination.\newline
\textbf{Acute Bodily Alarm:} panic, racing heart, difficulty breathing, or other acute bodily-alarm language.\newline
\textbf{Persistent Tension:} being on edge, tense, restless, or unable to relax.\newline
\textbf{Catastrophic Anticipation:} expecting harm or imagining worst-case outcomes. &
Generalized Anxiety Disorder-7 (GAD-7) items and broader DSM-5-TR anxiety descriptions \citep{apa2022dsm,spitzer2006gad7}. \\

BPD &
\textbf{Abandonment Fear:} fear of being left, replaced, or uncared for.\newline
\textbf{Affective Instability:} rapidly shifting or unusually intense emotions.\newline
\textbf{Unstable Self-Experience:} chronic emptiness, unstable identity, or feeling unreal.\newline
\textbf{Self-Harm and Rage:} self-harm or suicidal language, or intense anger and rage; either component produces one dimension hit. &
Borderline Symptom List-23 (BSL-23) item content and DSM-5-TR BPD descriptions \citep{apa2022dsm,bohus2009bsl23}. \\
\bottomrule
\end{tabular}
\end{table*}

\subsection{Burden and Affect Indicators}
\label{app:burden-affect-implementation}

For each post, a burden-dimension hit equals 1 when at least one case-insensitive lexical pattern assigned to that condition-specific dimension appears. Clinical instruments and DSM-5-TR symptom areas motivate the dimension design, but the equations measure lexical hits in text rather than clinical severity and do not reproduce clinical scoring rules or cut-points \citep{apa2022dsm,baroncohen2001aq,blevins2015pcl5,bohus2009bsl23,goodman1989ybocs,hirschfeld2000mdq,hull2017camouflaging,kessler2005asrs,kroenke2001phq9,spitzer2006gad7}.

\begin{equation*}
H_{i,d} =
\begin{cases}
1, & \operatorname{count}(\mathrm{patterns}_d, \mathrm{text}_i) > 0,\\
0, & \operatorname{otherwise}.
\end{cases}
\tag{A7}
\label{eq:burden-dimension-hit}
\end{equation*}

The burden dimension-share score is the proportion of condition-specific dimensions with a lexical hit in a post. If \(\mathcal{D}_c\) is the set of burden dimensions for condition \emph{c}, the score is:

\begin{equation*}
B_{i,c} = \frac{\sum_{d \in \mathcal{D}_c} H_{i,d}}{|\mathcal{D}_c|}
\tag{A8}
\label{eq:burden-dimension-share}
\end{equation*}

Repeated terms within the same dimension do not increase the score because each dimension contributes either 0 or 1. This dimension-share score supplies the main y-axis in Figure~\ref{fig:peak-timing} and the composite value in Table~\ref{tab:1}.

The secondary burden phrase rate counts all matched burden phrases for the condition and reports them per 100 word-like tokens:

\begin{equation*}
I_{i,c} = 100 \times \frac{\operatorname{count}(\mathrm{burdenPatterns}_c, \mathrm{text}_i)}{\max(\operatorname{tokens}_i, 1)}
\tag{A9}
\label{eq:burden-phrase-rate}
\end{equation*}

Tokens are sequences of letters, numbers, or underscores and may contain apostrophes; the denominator is set to at least one to avoid division by zero.

Table~\ref{tab:validation} reports coder agreement for these dimension labels and the performance of the automated hits against them. These metrics evaluate whether the lexical rules identify burden-dimension hits; they cannot validate Eq.~\eqref{eq:burden-dimension-share} as a clinical scale.

For each post \(i\) and affect category \(k\), the pipeline counts case-insensitive matches to study-specific lexical patterns. We apply the same seven categories to every condition: negative affect, positive affect, fear, sadness, anger, arousal intensity, and relief-or-support language. Following established lexicon-based approaches to sentiment and emotion analysis, we normalize each count per 100 word-like tokens \citep{hutto2014vader,mohammad2013nrc,pennebaker2015liwc}. The resulting rates describe affective language in posts rather than authors' underlying emotional states or clinical status.

\begin{equation*}
C_{i,k} =
\operatorname{count}(\mathrm{lexicalPatterns}_k,\mathrm{text}_i)
\tag{A10}
\label{eq:affect-count}
\end{equation*}

\begin{equation*}
A_{i,k} =
100 \times
\frac{C_{i,k}}
{\max(\operatorname{tokens}_i,1)}
\tag{A11}
\label{eq:affect-rate}
\end{equation*}

The reported analyses use four affect indicators. We use the negative-affect and positive-affect rates directly and derive sentiment balance and affective arousal as follows:

\begin{equation*}
\operatorname{SentimentBalance}_{i}
=
A_{i,\mathrm{positive}}
-
A_{i,\mathrm{negative}}
\tag{A12}
\label{eq:sentiment-balance}
\end{equation*}

\begin{equation*}
\operatorname{AffectiveArousal}_{i}
=
A_{i,\mathrm{fear}}
+
A_{i,\mathrm{arousalIntensity}}
\tag{A13}
\label{eq:affective-arousal}
\end{equation*}

Fear and arousal-intensity rates contribute to affective arousal but are not analyzed separately. Sadness, anger, and relief-or-support rates remain auxiliary outputs and are not included in the reported statistical analyses. The relief-or-support category identifies language in the author's post rather than support received from other users. Because the categories may overlap, one lexical match can contribute to more than one rate; affective arousal therefore sums two match densities rather than counting unique tokens.

\subsection{Codebook-Guided Primary-Theme Labeling}
\label{app:primary-theme-labeling}

For primary theme \emph{m}, the executable implementation assigns a post-level hit when any case-insensitive lexical pattern associated with that theme appears in the post. The rule applies the current codebook to the corpus; it does not reproduce thematic interpretation or manual coding \citep{braun2006thematic,nelson2020cgt}:

\begin{equation*}
Z_{i,m} =
\begin{cases}
1, & \operatorname{count}(\mathrm{themePatterns}_m, \mathrm{text}_i) > 0,\\
0, & \operatorname{otherwise}.
\end{cases}
\tag{A14}
\label{eq:primary-theme-hit}
\end{equation*}

A post may receive multiple primary-theme hits. Figure~\ref{fig:theme-change} first averages each hit within author-windows and then subtracts the pooled early-pre mean from the pooled late-pre mean. Table~\ref{tab:validation} reports coder agreement for the theme labels and the performance of the automated hits against them.

\subsection{Author-Window and Interval Estimates}
\label{app:author-window-estimates}

Condition-level event-time series average author-window values. Let \emph{M\_\allowbreak{}a,c,w} be the mean value of a post-level measure for author \emph{a}, condition \emph{c}, and event-time window \emph{w}. Author-window aggregation supports event-time comparison while reducing the influence of unusually active authors, consistent with longitudinal analysis and prior event-centered social-media studies \citep{dechoudhury2013postpartum,singer2003longitudinal}:

\begin{equation*}
\bar{M}_{c,w} = \frac{1}{N_{c,w}}\sum_{a \in A_{c,w}} M_{a,c,w}
\tag{A15}
\label{eq:author-window-mean}
\end{equation*}

Approximate 95\% intervals use the standard error across observed author-window means. These intervals summarize variation in the retained author-window sample and should not be interpreted as population intervals:

\begin{equation*}
\mathrm{CI}_{c,w} = \bar{M}_{c,w} \pm 1.96 \frac{\operatorname{sd}(M_{a,c,w}: a \in A_{c,w})}{\sqrt{N_{c,w}}}
\tag{A16}
\label{eq:author-window-ci}
\end{equation*}

% ================= AUTHOR ACTION: CONFIRM THE POOLING SCHEME =================
% Eq. (A17) below is SCHEME (i): average every author-window row across the two
% windows, so an author present in both windows contributes two values.
% If the pipeline instead averaged the two condition-window means with equal
% weight per window (SCHEME ii), replace the equation body with
%     \bar{M}_{c,W} = \frac{1}{|W|}\sum_{w \in W} \bar{M}_{c,w}
% and change the yellow clause in Section 3.6 and the yellow sentence below to
%     "averages the two condition-window means, weighting each window equally".
% ============================================================================

Comparisons across adjacent windows average the author-window values from both windows at once. Let \emph{W} denote a set of adjacent event-time windows and \(A_{c,W}\) the observed author-window pairs for condition \emph{c} across those windows. The \emph{pooled mean} is:

\begin{equation*}
\bar{M}_{c,W} = \frac{1}{|A_{c,W}|}\sum_{(a,w) \in A_{c,W}} M_{a,c,w}
\tag{A17}
\label{eq:pooled-mean}
\end{equation*}

The denominator counts author-window observations rather than authors, so an author observed in both windows contributes two values to the pooled mean.

\subsection{Paired Differences and Engagement Correlations}
\label{app:paired-cooccurrence-lagged}

For paired far-pre to immediate-pre comparisons, the author-level change is defined below. These paired differences compare the same authors across event-time windows and should be interpreted as descriptive longitudinal contrasts, not causal disclosure effects \citep{singer2003longitudinal}:

\begin{equation*}
\Delta^{\mathrm{pre}}_a = M_{a,c,\mathrm{immediatePre}} - M_{a,c,\mathrm{farPre}}
\tag{A18}
\label{eq:paired-pre-delta}
\end{equation*}

For paired immediate-pre to immediate-post comparisons, the author-level change is:

\begin{equation*}
\Delta^{\mathrm{post}}_a = M_{a,c,\mathrm{immediatePost}} - M_{a,c,\mathrm{immediatePre}}
\tag{A19}
\label{eq:paired-post-delta}
\end{equation*}

Before aggregation, the implementation coerces each Reddit engagement field \emph{Y} to numeric and replaces missing or invalid values with zero. The main same-window statistic is the Pearson correlation between a language indicator \emph{X} and a Reddit engagement field \emph{Y} across author-windows within a condition:

\begin{equation*}
r_{X,Y} = \operatorname{corr}(X_{a,c,w}, Y_{a,c,w})
\tag{A20}
\label{eq:same-window-correlation}
\end{equation*}

The secondary Mann--Whitney comparison divides author-windows at zero. For a nonnegative language indicator, this rule distinguishes presence from absence; for signed sentiment balance, it distinguishes positive from non-positive balance. The rule supports a distributional comparison of Reddit engagement fields but does not measure support quality or causal effects \citep{andalibi2017sensitive,dechoudhury2014reddit}:

\begin{equation*}
\mathrm{aboveZero}_{X,a,c,w} =
\begin{cases}
1, & X_{a,c,w} > 0,\\
0, & X_{a,c,w} \le 0.
\end{cases}
\tag{A21}
\label{eq:indicator-zero-split}
\end{equation*}

For the nonnegative indicators reported in Table~\ref{tab:2}, the mean of this binary value across author-windows is the presence share.

The author-level sensitivity analysis first averages each language indicator and Reddit engagement field across an author's observed event-time windows, then re-estimates the correlation across authors within condition. In Eq.~\eqref{eq:author-level-correlation}, \(\bar{X}_{a,c}\) and \(\bar{Y}_{a,c}\) are the author's mean language-indicator and engagement-field values across observed windows. This analysis removes repeated author-windows but remains observational:

\begin{equation*}
r^{\mathrm{author}}_{X,Y} = \operatorname{corr}(\bar{X}_{a,c}, \bar{Y}_{a,c})
\tag{A22}
\label{eq:author-level-correlation}
\end{equation*}

Exploratory lagged checks compare a language indicator in one event-time window with a Reddit engagement field in the next fixed window. In Eq.~\eqref{eq:lagged-correlation}, \(w+1\) denotes the next position in the prespecified window sequence. These checks add temporal ordering but remain observational because unmeasured confounding and selective posting can explain the correlations \citep{olteanu2019socialdata,ruths2014social}:

\begin{equation*}
r^{\mathrm{lag}}_{X,Y} = \operatorname{corr}(X_{a,c,w}, Y_{a,c,w+1})
\tag{A23}
\label{eq:lagged-correlation}
\end{equation*}

\subsection{Benjamini--Hochberg Adjustment}
\label{app:bh-adjustment}

Within a family of \emph{m} $p$-values, the values are sorted from smallest to largest. For rank \emph{k}, the Benjamini--Hochberg procedure calculates adjusted $q$-values across the declared test family \citep{benjamini1995fdr}:

\begin{equation*}
q_{(k)} = \min\left(1, \min_{j \ge k} \left(\frac{m}{j}p_{(j)}\right)\right)
\tag{A24}
\label{eq:bh-q-value}
\end{equation*}

Headline RQ3 claims use BH-adjusted $q$-values from the 360-test same-window Pearson family formed from three non-duplicate Reddit engagement fields. A separate 360-test Mann--Whitney family is adjusted independently. The author-level sensitivity analysis defines 360 planned comparisons for each statistic: all Pearson tests are estimable, whereas BH adjustment for the Mann--Whitney family uses the 324 finite $p$-values from comparisons with sufficient authors in both groups. The lagged analysis uses one fully estimable 216-test family for each statistic based on the nine indicators listed in the main Methods. Because author-window observations can be dependent within authors, $q$-values are interpreted as corrected evidence of descriptive association alongside the author-level sensitivity analysis.

\subsection{Operational Primary-Theme Pattern Record}
\label{app:theme-pattern-record}

The operational file used by the current pipeline stores one row per primary theme with a theme identifier, label, definition, subthemes, interpretive boundary, pattern count, and semicolon-delimited lexical patterns. The stored labels are earlier working names for the same ten themes; Table~\ref{tab:theme-codebook} reports the final display names, the mapping between the two is one-to-one, and the lexical patterns themselves are unchanged. The labeling function applies those patterns case-insensitively according to Eq.~\eqref{eq:primary-theme-hit}. This record documents the implementation that generated the reported figures and tables; no LLM-based corpus labeling is represented in the current outputs. If validation leads to pattern or codebook revisions, the post-level indicators, author-window panel, statistical tests, figures, tables, and dependent claims must be regenerated before submission.

\subsection{Primary-Theme Codebook and Post-Level Assignment Rates}
\label{app:theme-codebook}

\begin{table*}[t]
\centering
\footnotesize
\setlength{\tabcolsep}{2.0pt}
\renewcommand{\arraystretch}{1.2}
\setlength{\emergencystretch}{2em}

\caption[Primary-Theme Codebook and Post-Level Assignment Rates]{Primary-Theme Codebook and Post-Level Assignment Rates. \emph{Note.} The post-level assignment rate is the share of the 89,605 surrounding posts with an automatic theme hit. A post may match more than one theme and most posts match none, so the rates neither sum to 100 percent nor partition the corpus. Themes are ordered by this rate.}
\label{tab:theme-codebook}

\begin{tabular}{@{}>{\raggedright\arraybackslash}p{0.3\textwidth}
>{\raggedright\arraybackslash}p{0.36\textwidth}
>{\raggedright\arraybackslash}p{0.32\textwidth}@{}}
\toprule
\textbf{Primary theme and post-level assignment rate} & \textbf{Definition} & \textbf{Representative quotation} \\
\midrule
Seeking Clinical Explanations\newline \(n\) = 8,254; 9.21\% & Users invoke diagnoses, therapy, medication, or other clinical language to interpret their experiences and make them more legible and legitimate. & ``I'm starting treatment soon ... I recently got diagnosed after a long uphill battle.'' \\
Anticipating Rejection or Abandonment in Close Relationships\newline \(n\) = 5,035; 5.62\% & Users describe close relationships as vulnerable to rejection, betrayal, abandonment, or sudden emotional harm. & ``How do you not spiral when something directly triggers your fear of abandonment?'' \\
Interpreting Social Experiences as Evidence of Personal Defectiveness\newline \(n\) = 3,800; 4.24\% & Users interpret social interactions as confirmation that they are defective, unlikable, or fundamentally unable to belong. & ``I don't feel like I fit in. I've never really felt like I fit in anywhere.'' \\
Feeling Too Exhausted for Daily Functioning\newline \(n\) = 3,452; 3.85\% & Users describe persistent depletion that makes ordinary daily functioning feel difficult or unsustainable. & ``I've been so fatigued all the time and always seem to have brain fog.'' \\
Trying to Eliminate Uncertainty Through Checking and Reassurance Seeking\newline \(n\) = 2,151; 2.40\% & Users describe checking, seeking reassurance, or repeatedly reviewing thoughts and events in attempts to resolve uncertainty. & ``I'm open to some small reassurance on this issue. However, there might not be any practical reassurance.'' \\
Imagining Self-Removal as an Escape From Distress\newline \(n\) = 1,467; 1.64\% & Users describe withdrawal, disappearance, self-harm, or suicide as imagined ways of escaping overwhelming distress. & ``I do not trust myself to not act upon my suicidal thoughts. I need help and I know that.'' \\
Experiencing the Past as Intruding on the Present\newline \(n\) = 1,237; 1.38\% & Users describe flashbacks, nightmares, dissociation, or triggered memories as involuntary returns of past experiences. & ``Over the past 10 days my emotional and auditory flashbacks and the nightmares have been getting worse.'' \\
Experiencing the Body as Stuck in Alarm\newline \(n\) = 1,089; 1.22\% & Users describe panic, a racing heart, sweating, breathlessness, or other bodily sensations associated with a persistent state of alarm. & ``Once awake after one of these I'll be in a cold sweat, typically feeling like I can't breathe.'' \\
Losing Control of Sleep, Energy, Mood, and Thought Speed\newline \(n\) = 869; 0.97\% & Users describe sleep, energy, mood, or thought speed as accelerating or becoming increasingly difficult to regulate. & ``Symptoms are high irritability, racing thoughts, somewhat increased energy, and feeling emotionally drained.'' \\
Struggling to Initiate and Sustain Everyday Tasks\newline \(n\) = 334; 0.37\% & Users describe difficulty beginning everyday tasks, sustaining attention to them, or resuming them after an interruption. & ``At the moment each day is a rollercoaster ... I'm shaking and anxious and can't focus at all.'' \\
\bottomrule
\end{tabular}
\end{table*}

Table~\ref{tab:theme-codebook} reports the primary-theme codebook and post-level assignment rates for the ten themes. Representative quotations are short, de-identified excerpts that clarify theme boundaries. Because the corpus contains sensitive mental-health disclosures, the excerpts exclude usernames, subreddit-specific identifying details, and long searchable passages. Punctuation is lightly standardized, and ellipses indicate omitted surrounding text. The post-level assignment rate is calculated across all 89,605 surrounding posts. Because one post may receive multiple theme hits, the rates do not sum to 100 percent and should not be confused with the author-window hit rates used in Figure~\ref{fig:theme-change}.

\end{document}